\documentclass[sn-mathphys,Numbered]{sn-jnl-mod}

\usepackage{graphicx}%
\usepackage{multirow}%
\usepackage{amsmath,amssymb,amsfonts}%
\usepackage{amsthm}%
\usepackage{mathrsfs}%
\usepackage[title]{appendix}%
\usepackage{xcolor}%
\usepackage{textcomp}%
\usepackage{manyfoot}%
\usepackage{booktabs}%
\usepackage{algorithm}%
\usepackage{algorithmicx}%
\usepackage{algpseudocode}%
\usepackage{listings}%

\definecolor{viol}{rgb}{0.7, 0.4, 1}

\usepackage{placeins}
\usepackage{ulem}
\usepackage{soul}

\usepackage{hyperref}

\definecolor{mygray}{cmyk}{0, 0, 0, 0.3}

\usepackage{float}

\begin{document}

\title[Article Title]{Electron Spin Resonance Spectroscopy on Magnetic Van der Waals Compounds}


\author*[1]{\fnm{Vladislav} \sur{Kataev}}\email{v.kataev@ifw-dresden.de}

\author[1,2]{\fnm{Bernd} \sur{B\"uchner}}\email{b.buechner@ifw-dresden.de}

\author[1]{\fnm{Alexey} \sur{Alfonsov}}\email{a.alfonsov@ifw-dresden.de}

\affil*[1]{\orgname{Leibniz Institute for Solid State and Materials Research Dresden}, \orgaddress{\street{Helmholtzstr. 20}, \city{Dresden}, \postcode{D-01069}, \country{Germany}}}

\affil[2]{\orgname{Institute for Solid State and Materials Physics and W{\"u}rzburg-Dresden Cluster of Excellence ct.qmat, TU Dresden}, \city{Dresden}, \postcode{D-01062}, \country{Germany}}



\abstract{The field of research on magnetic van der Waals compounds -- a special subclass of quasi-two-dimensional materials -- is currently rapidly expanding due to the relevance of these compounds to fundamental research where they serve as a playground for the investigation of different models of quantum magnetism and also in view of their unique magneto-electronic and magneto-optical properties pertinent to novel technological applications. The Electron Spin Resonance (ESR) spectroscopy plays an important role in the exploration of the rich magnetic behavior of van der Waals compounds due to its high sensitivity to magnetic anisotropies and unprecedentedly high energy resolution that altogether enable one to obtain thorough insights into the details of the spin structure in the magnetically ordered state and the low-energy spin dynamics in the ordered and paramagnetic phases. This article provides an overview of the recent achievements in this field made by the ESR spectroscopic techniques encompassing representatives of antiferro- and ferromagnetic van der Waals compounds of different crystal structures and chemical composition as well as of a special category of these materials termed magnetic topological insulators.    }


\maketitle

\section{Introduction}\label{sec:Introduction}

The successful isolation of graphene in 2004 \cite{Novoselov2004} -- a monolayer of carbon exfoliated from bulk graphite -- has opened a vast new field of research on two-dimensional (2D) materials which, besides graphene and its analogs, include by now III-VI semiconductors, e.g., GaSe, InSe, In$_2$Se$_3$, transition-metal dichalcogenides, e.g., MoS$_2$, WS$_2$, TiS$_2$ RbSe$_2$, MoTe$_2$, and oxides, e.g., TiO$_2$, MoO$_3$, WO$_3$ \cite{Geim2013,Ajayan2016}. Most of these compounds are in the bulk form layered van der Waals crystals which by virtue of the weak interlayer van der Waals forces can be relatively easy thinned by exfoliation. In the 2D limit they feature outstanding physico-chemical, electronic and optical properties often distinct from the bulk  that make these materials extremely interesting for novel technological applications. More recently the focus of the 2D research was extended to magnetic van der Waals compounds promising new magneto-electronic and magneto-optical functionalities (for recent reviews see, e.g., Refs.~\cite{Gibertini2019,Yang2021,Xu2022,Liu2023}).    

Van der Waals magnets appear to be attractive both as model systems for investigations of  fundamental aspects of magnetism in reduced spatial dimensions and also regarding the practical use in the next-generation spintronic devices. Though many of them were known already since long time and their elementary magnetic properties were reported in the literature, magnetic studies of these compounds experience currently a renaissance due to improvements in the material synthesis and the progress of the measurement techniques that make possible to address new scientific questions arising in the context of the 2D physics. 

Among the employed experimental methods the Electron Spin Resonance (ESR) spectroscopy and related ferromagnetic (FM) and antiferromagnetic (AFM) resonance techniques, FMR and AFMR, play an important role in obtaining valuable insights into the static and dynamic magnetic properties of van der Waals compounds. This became possible due to enormous instrumental development since the discovery of ESR 80 years ago in terms of the sensitivity, expansion of the frequency range towards far-infrared region and the use of very strong magnetic fields. Still studies of magnetic resonance on the monolayers remain very challenging.

 Fortunately, in the absence of the strong covalent bonds between the layers, which could provide the pathway for the conventional superexchange interaction \cite{Anderson1959}, the interlayer magnetic coupling in the van der Waals magnets is very weak. Instead theories suggest that the interlayer coupling originates from the weak overlap of the ligands' $p$-electron densities between the layers giving rise to a kind of super-superexchange between the magnetic ions in the nighboring planes that is by orders of magnitude smaller that the intraplane exchange \cite{Sivadas2018,Jang2019}. This implies that important clues about the magnetism of the monolayers can be obtained from the measurements on bulk crystals which  due to such very weak interlayer coupling are naturally quasi-2D magnets and as such may demonstrate inherent 2D magnetic phenomena. 

%

This review focuses on the results of magnetic resonance investigations mainly on the bulk antiferromagnetic van der Waals compounds in the magnetically ordered and paramagnetic states and to a less extent on their ferromagnetic counterparts, the FMR studies of which were summarized in a recent review in Ref.~\cite{Tang2023}.

\section{Antiferromagnetic van der Waals compounds}\label{sec:AFM_vdW}	

AFM van der Waals compounds are of a significant interest both for fundamental and applied research due to their layered structure with relatively weak intraplane and even much weaker interplane magnetic interactions as compared  to, e.g., magnetic oxides. This significantly reduces the N\'{e}el ordering temperature $T_{\rm N}$ and  makes it possible to influence their magnetic state by applying moderate magnetic fields and probe magnetic excitations by GHz and sub-THz resonance techniques. From the fundamental perspective, in van der Waals antiferromagnets different spin models can be realized  based on the AFM anisotropic Heisenberg, Ising-type, or planar $XY$ Hamiltonians, and their properties can be studied and compared with theoretical predictions. Regarding applications, these materials are interesting in view of the antiferromagnetic spintronic devices featuring ultrafast spin dynamics and being insensitive to disturbing magnetic fields \cite{Jungwirth2016,Baltz2018}. For these reasons a significant number of AFM van der Waals compounds were invstigated with ESR and AFMR spectroscopies. An overview of the results of these studies will be presented in the following sections.

\subsection{CrCl$_3$}\label{subsec:CrCl3}

	\begin{figure}
	\centering
	\includegraphics[width=\linewidth]{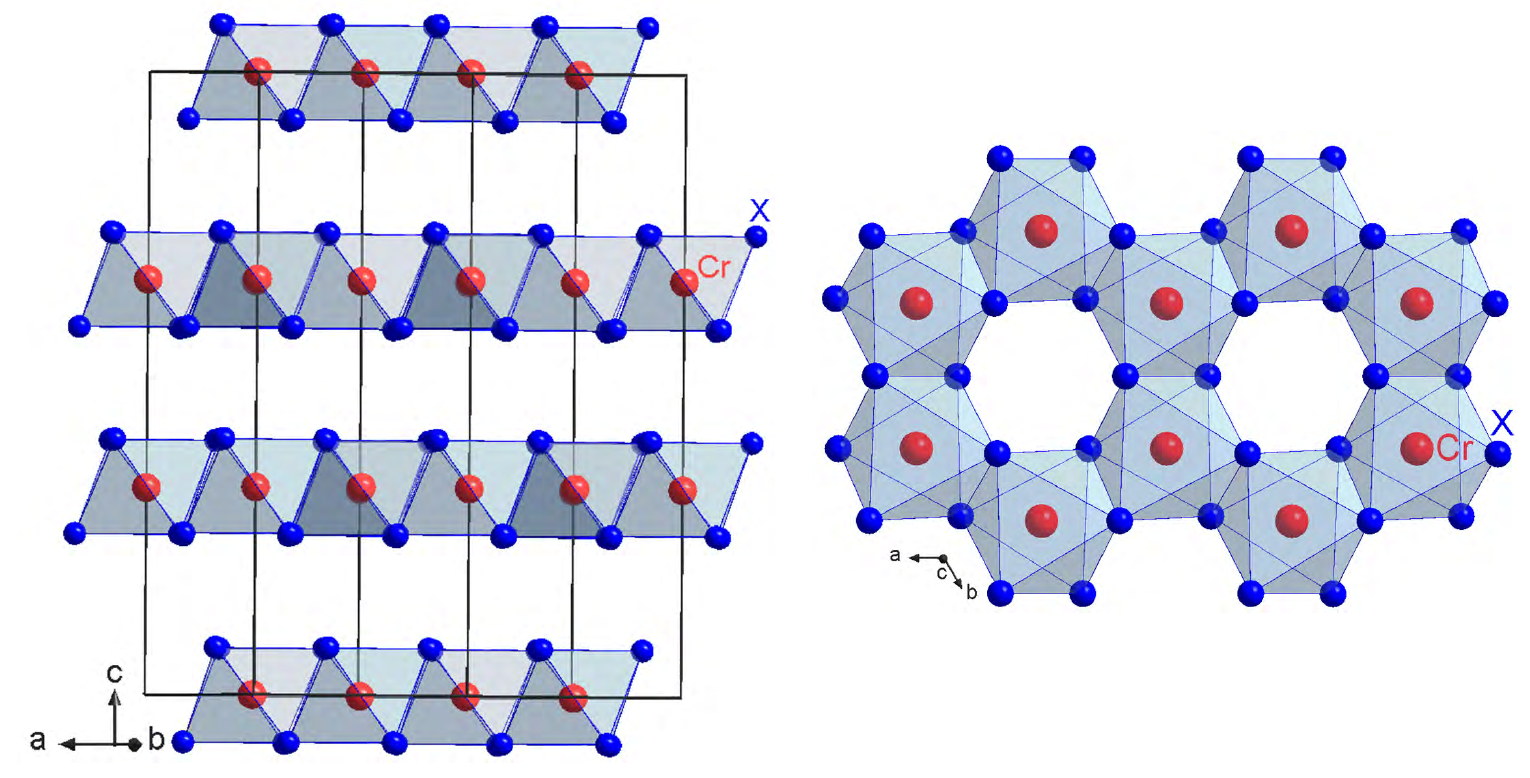}
	\caption{Crystal structure of the chromium-based trihalides viewed along the $ab$-plane (left) and along the $c$-axis (right). Cr$X_6$ ($X$ = Cl, Br, I) octahedra are arranged on a honeycomb plane, the planes are stacked along the $c$-axis.}
	\label{fig:structure_CrX3}
\end{figure}

In CrCl$_3$, the Cr$^{3+}$ ($3d^3,\, S = 3/2,\, L = 0)$ ion is covalently bonded to six Cl$^{-}$ ligands forming a regular CrCl$_6$ octahedron. These octahedra share edges and form a 2D-honeycomb lattice in the $ab$-crystal plane (Fig.~\ref{fig:structure_CrX3}). The Cr spins in the $ab$-plane are coupled ferromagnetically with the nearest-neighbor exchange constant $J^{\rm ab} = -5.25$\,K (-0.45 meV) \cite{Narath1965} and undergo a 2D-FM transition at $T_{\rm c}^{\rm 2D} \approx17$\,K with the spins lying in the plane \cite{Kuhlow1982,Bastien2019}. 
 A large spatial separation between the honeycomb planes along the $c$-axis amounts to about 5.81\,\AA\ and results in a weak antiferromagnetic inter-layer coupling $J^{\rm c} = 18$\,mK (1.55\,$\mu$V) \cite{Narath1965} that yields a 3D-AFM order at $T_{\rm N} \approx 15$\,K \cite{Kuhlow1982} (14\,K in Ref.~\cite{Bastien2019}).
%

ESR studies of CrCl$_3$ have a long history tracing back to the early days of ESR. It was one of the first compounds whose resonance signal was reported by Evgenii Zavoisky soon after his discovery of electron paramagnetic resonance \cite{Zavoisky1946,Kochelaev1995}. The interest in this material was much later renewed in the context of physics of low-dimensional spin systems and functional magnetic properties of van der Waals magnets. Already in 1991 \citet{Chehab1991} pointed out the prominence of the specific 2D spin dynamics far above the magnetic ordering temperature of CrCl$_3$. It is characterized by the dominance of the long-wave modes (wave vector $q\rightarrow 0$) of spin fluctuations manifesting in the $(3\cos^2\theta - 1)^2$ type of the angular dependence of the ESR linewidth \cite{Benner1990}. It is distinct from the case of 3D spin systems where fluctuations with all wave vectors are present giving rise to the $(\cos^2\theta + 1)$ type of the angular behavior \cite{Benner1990}.       

Magnetic excitations in the 3D-AFM ordered state of CrCl$_3$ were studied by broad-band AFMR spectroscopy in Ref.~\cite{MacNeil2019} for the external magnetic field applied in the $ab$-crystal plane. Since CrCl$_3$ in the ground state can be viewed as a two-sublattice collinear antiferromagnet, two types of AFMR modes can be expected. The first one, the so-called acoustic mode, corresponds to the in-phase precession of the magnetizations of the AFM-coupled neighboring FM-ordered Cr planes (Fig.~\ref{fig:structure_CrX3}) \cite{MacNeil2019}:  

\begin{equation}
	\omega_- = \mu_0\gamma \sqrt{2H_{\rm e}(2H_{\rm e} + M_{\rm  s})}\frac{H}{2H_{\rm e}}\ . \label{eq:acoustic_mode}
\end{equation}

The second, optical mode, is the anti-phase precession of the respective magnetizations \cite{MacNeil2019}:

\begin{equation}
	\omega_+ = \mu_0\gamma \sqrt{2H_{\rm e}M_{\rm s}(1-\frac{H^2}{4H_{\rm e}^2})}\ . \label{eq:optical_mode}
\end{equation}

Here, $\mu_0$ and $\gamma$ are the vacuum permeability and the gyromagnetic ratio, and $H_{\rm e}$ and $M_{\rm s}$ are the exchange field and the saturation magnetization, respectively.

	\begin{figure}
	\centering
	\includegraphics[width=\linewidth]{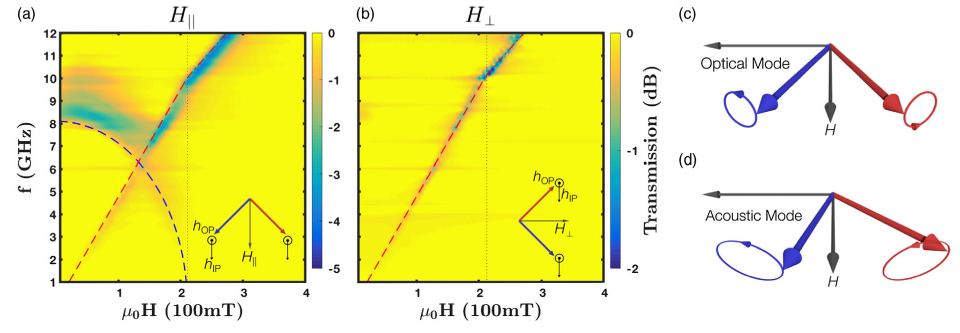}
	\caption{Frequency vs. magnetic field diagram of the AFMR modes in CrCl$_3$ for the external field $\mathbf{H}$ applied in the $ab$ crystal plane parallel to the magnetic component of the microwave field $\mathbf{H}\parallel \mathbf{H_{mw}}$ ($H_\parallel$) (a), and perpendicular to it $\mathbf{H} \perp \mathbf{H_{mw}}$ ($H_\perp$)  (b). For the $H_\parallel$ geometry both optical and acoustic AFMR modes were observed, whereas for $H_\perp$ only acoustic mode is visible. Red and blue dashed lines in both panels are fits of the acoustic and optical mode frequencies to Eqs.~(\ref{eq:acoustic_mode}) and (\ref{eq:optical_mode}), respectively. Precession of the sublattice magnetizations in the optical and acoustic mode are schematically illustrated in panels (c) and (d), respectively. (Reprinted with permission from D.~MacNeill {\it et al.}, Phys. Rev. Lett. {\bf 123} (4), 047204 (2019) \cite{MacNeil2019}. Copyright (2019) by the American Physical Society.)}
	\label{fig:MacNeil2019}
\end{figure}

Owing to the expected smallness of the interlayer magnetic exchange of the Cr $ab$-planes \cite{Narath1965} 
%
%
both modes could be detected at relatively low frequencies $f < 12$\,GHz (Fig.~\ref{fig:MacNeil2019}.) Indeed, fitting of the obtained $f(H)$ frequency dependences to Eqs.~(\ref{eq:acoustic_mode}) and (\ref{eq:optical_mode}) [Figs.~\ref{fig:MacNeil2019}(a) and (b)] yielded a small value of the exchange field $\mu_0H_{\rm e} = 105$\,mT. Consequently, for applied fields $H>2H_{\rm e}$ the Cr spins become fully polarized along the field direction and the magnetic system of CrCl$_3$ enters the forced ferromagnetic state. Similar observation of the optical AFM mode in a CrCl$_3$ crystal coupled to the superconducting cavity was recently reported in Ref.~\cite{Zhang2021}.

The dynamic properties of the spin-polarized state in CrCl$_3$ were addressed in Ref.~\cite{Zeisner2020} by high-field ESR spectroscopy covering the frequency range up to 350\,GHz. In this case the FMR signal of the Cr spins forcefully aligned by the external field $\mathbf{H}$ appears anisotropic, shifting differently from the expected position of the high-temperature paramagnetic resonance signal for $\mathbf{H}$ applied parallel and normal to the crystal $c$-axis. The measured field dependence of the FMR modes for $\mathbf{H}\parallel \mathbf{c}$ and  $\mathbf{H}\perp \mathbf{c}$ configurations were analyzed in the framework of a conventional FMR theory \cite{Smit1955,Skrotskii1966,Farle1998}. In agreement with previous studies \cite{McGuire2017,Bastien2019,MacNeil2019,Bizette1961} it has been shown that the observed anisotropy is essentially due to the platelet-like shape of the measured crystal whereas the intrinsic, magnetocrystalline anisotropy is practically negligible. This enables one to classify CrCl$_3$ as an example of an almost ideal isotropic Heisenberg magnet. 

It should be noted that even far above the ordering temperature $T_{\rm N} \approx 15$\,K the anisotropic temperature-dependent shifts of the ESR signal measured at high frequencies such as 90\,GHz in Ref.~\cite{Zeisner2020} and 120\,GHz in Ref.~\cite{Saiz2019} are still present up to $\sim 3T_{\rm N}$ suggesting that the corresponding strong magnetic field promotes in-plane ferromagnetic correlations even in the paramagnetic state, which appears to be a salient feature of a quasi-2D van der Waals magnet.  

Interestingly, Ref.~\cite{Singamaneni2020} reported that exposure of single crystals of CrCl$_3$ to visible light affects the parameters of the ESR signal of Cr$^{3+}$ ($3d^3,\, S = 3/2$) ions measured at the X-band frequency of 9.45\,GHz , their intensity, $g$-factor, and the linewidth. This sensitivity to light was ascribed to the photo-induced electron transitions between the valence band and localized Cr$^{2+}$ ($3d^4,\, S = 2$) levels. The similar effect was observed by the same authors for the ferromagnetic CrI$_3$ (see Sect.~\ref{subsec:CrX3}) suggesting a possible light-control functionality of these compounds in the monolayer limit.

Besides observing the above discussed uniform acoustic and optical resonance modes \citet{Kapoor2021} were able to detect in a 20\,$\mu$m thin single crystal CrCl$_3$ additional non-uniform modes which were ascribed to the standing spin waves excited across the thickness of a crystal. They manifest themselves as additional multiple resonances in the vicinity of the uniform modes. The distance of each spin-wave resonance from the respective uniform mode scales linearly with its number as predicted by the theory of standing spin waves in a material with a nonuniform volume magnetization \cite{Portis1963}. The authors could estimate the lifetime of the resonantly excited magnons to be of the order of a few ns. Since spin waves can carry information without involving the electron charge they promise to be an efficient means of low-loss data transfer in spintronic devices where, as the authors believe, a few-layer CrCl$_3$ can be used as a suitable medium.

\subsection{The family of $M_2$P$_2$S$_6$ }\label{subsec:MPS}

	\begin{figure}
	\centering
	\includegraphics[width=\linewidth]{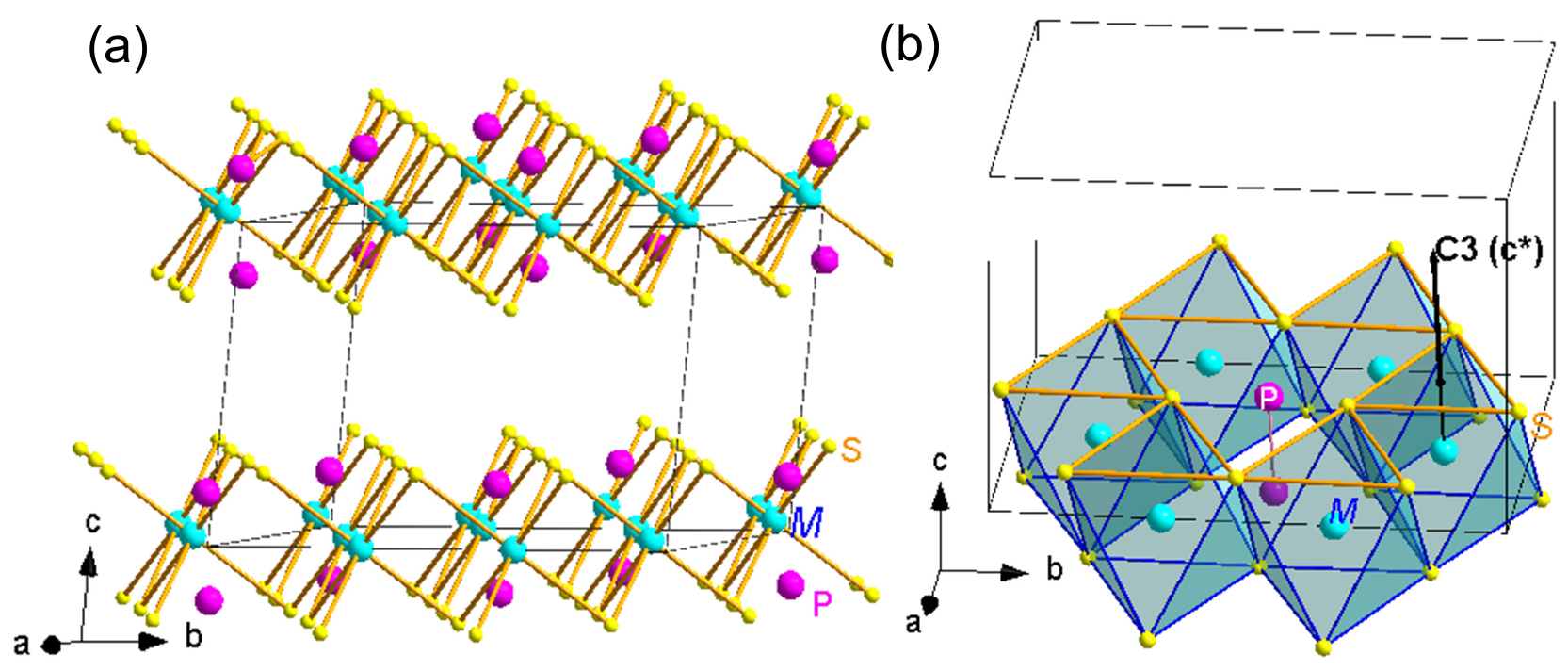}
	\caption{(a) Crystal structure of the $M_2$P$_2$S$_6$ ($M$ = Mn, Fe, Ni, Mn$_{0.5}$Ni$_{0.5}$, Cu$_{0.5}$Cr$_{0.5}$) compounds of the monoclinic symmetry, space group $C2/m$. The atoms are colored as $M$ = turquoise, P = pink, and S = yellow. (b) Individual $M_2$P$_2$S$_6$ layer. The $M$S$_6$ octahedra form a honeycomb lattice with the $C_3$ symmetry axis of the octahedra normal to the $ab$-plane ($c^*$-axis).  The P-P dumbbell occupies the void of each honeycomb. } 
	\label{fig:structure_M2P2S6}
\end{figure}

The family of metal tiosulfates $M_2$P$_2$S$_6$ ($M$ = Mn, Fe, Ni, Mn$_{0.5}$Ni$_{0.5}$, Cu$_{0.5}$Cr$_{0.5}$) represents an interesting subclass of van der Waals magnets where the transition metal ions with different spin multiplicities and single ion anisotropies can occupy the $M$-site in the crystal structure  (Fig.~\ref{fig:structure_M2P2S6}). The metal ions are 6-fold coordinated by sulfur ligands and are arranged on a honeycomb lattice [Fig.~\ref{fig:structure_M2P2S6}(b)]. The intra-plane exchange interaction between the $M$ spins is antiferromagnetic and sizable whereas the coupling between the planes is weak due to a large spatial interplane separation
%
%
Thus, the $M_2$P$_2$S$_6$ family offers realization of various quasi-2D AFM spin models ranging from the Ising and Heisenberg to the $XY$ Hamiltonians. In the following, ESR works on most frequently studied members of this family with $M$ = Mn, Fe, Ni, Mn$_{0.5}$Ni$_{0.5}$, and Cu$_{0.5}$Cr$_{0.5}$ will be briefly reviewed.           

\subsubsection{Ni$_2$P$_2$S$_6$}
\label{subsubsec:NPS}

Ni$_2$P$_2$S$_6$ is an antiferromagnet ordering at $T_{\rm N} = 155$\,K due to residual, weak inter-plane coupling. It has been classified as an "easy"-plane antiferromagnet, i.e., the ordered spins in the honeycomb lattice lie in the plane where a preferred spin orientation along an in-plane "easy"-axis was identified \cite{Joy1992,Selter2021,Wildes2015,Lancon2018}. The Ni$^{2+}$ ion in Ni$_2$P$_2$S$_6$ possesses 8 electrons on the 3$d$ shell and carries a spin $S = 1$. In the presence of a uniaxial distortion of the surrounding ligand electrical field the spin triplet splits into a singlet  $\mid0\rangle$ and a doublet $\mid\pm 1\rangle$ giving rise to a single-ion anisotropy (SIA) of the easy-axis or the easy-plane type depending on whether the spin doublet or the spin singlet have the lowest energy \cite{AbragamBleaney2012}. Such a distortion along the $C_3$ symmetry axis of the NiS$_6$ octahedron is present in Ni$_2$P$_2$S$_6$ [Fig.~\ref{fig:structure_M2P2S6}(b)] and the related single ion anistropy can potentially contribute to the total anisotropy of the AFM-ordered Ni spin lattice. The strength and the sighn of SIA parameter $D$ can be deduced from the anisotropy of the $g$-factor. Given that the orbital moment of the Ni$^{2+}$ ion is quenched in fist order, the $g$-anisotropy is expected to be small and can be resolved only by high-field ESR spectroscopy owing to its excellent spectral resolution.  
 
\begin{figure}
	\centering
	\includegraphics[clip,width=\columnwidth]{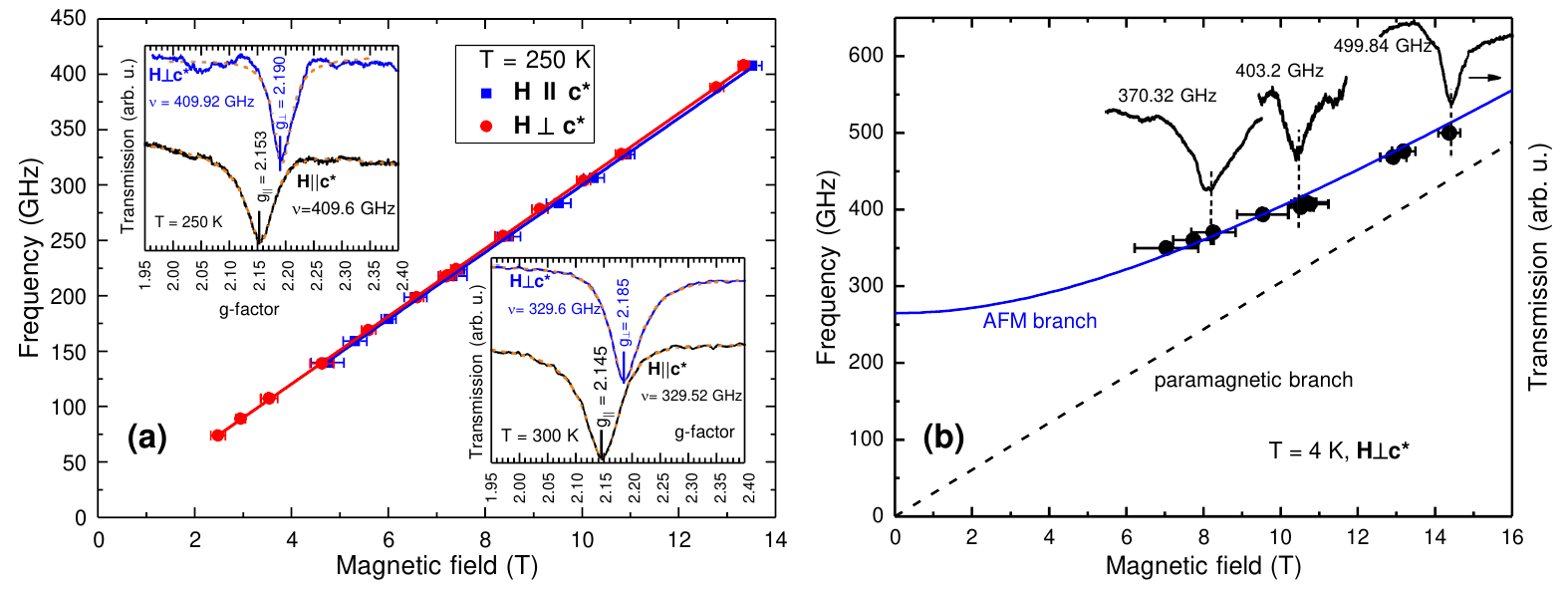}
	\caption{High-field ESR results on Ni$_2$P$_2$S$_6$. (a) $\nu(H_{\rm res})$ dependence of the ESR signals at $T = 250$\,K for ${\bf H}\parallel {\bf c^\ast}$ (blue squares) and ${\bf H}\perp {\bf c^\ast}$ (red circles). Solid lines represent the fit according to the resonance condition $h\nu = g\mu_{\rm B}\mu_0H_{\rm res}$. Insets: HF-ESR signals (black and blue solid lines) for ${\bf H}\parallel {\bf c^\ast}$ and ${\bf H}\perp {\bf c^\ast}$ plotted as a function of the $g$-factor $ g = h\nu/\mu_{\rm B}\mu_0H$  at $\nu \approx 410$\,GHz and $T = 300$\,K (top left) and $\nu \approx 330$\,GHz and $T = 250$\,K (bottom right). Orange dash lines are fits to the Lorenzian line profile. The $g$-values of the resonance peaks are indicated in the respective plots. (b) Left vertical scale: $\nu(H_{\rm res})$ dependence of the ESR signal at $T = 4$\,K for ${\bf H}\perp {\bf c^\ast}$ (circles). Solid line depicts the fit of the data to 
		the easy-plane AFM resonance branch fit function $\nu =h^{-1}[(g_{\perp}\mu_{\rm B}\mu_0H_{\rm res})^2+\Delta^2]^{1/2}$. 
		Dashed line corresponds to the paramagnetic resonance condition $\nu = h^{-1}g_{\perp}\mu_{\rm B}\mu_0H_{\rm res}$. Right vertical scale: HF-ESR signals at selected frequencies. (Reprinted with permission from K.~Mehlawat {\it et al.}, Phys. Rev. B. {\bf 105} (21), 214427 (2022) \cite{Mehlawat2022}. Copyright (2022) by the American Physical Society.)} 
	\label{fig:ESR_Ni2P2S6}
\end{figure}

This has been shown in Ref.~\cite{Mehlawat2022} reporting the results of the ESR measurements on single crystals of Ni$_2$P$_2$S$_6$ in the broad frequency and magnetic field range in the paramagnetic and AFM-ordered states. In the paramagnetic regime far above $T_{\rm N} = 155$\,K the Ni$^{2+}$ ESR signal is almost isotropic. If measured at frequencies $\nu < 150$\,GHz the resonance field $H_{\rm res}$ appears apparently  independent from whether the external field is applied in the $ab$-plane or normal to it [Fig.~\ref{fig:ESR_Ni2P2S6}(a), main panel]. However, increasing the frequency up to 410\,GHz enabled one to clearly resolve the anisotropy of the ESR signal [Fig.~\ref{fig:ESR_Ni2P2S6}(a), insets] and quantify the $g$-factor tensor elements $g_{\parallel} = 2.149 \pm 0.004$ and $g_{\perp} = 2.188 \pm 0.004$. Based on this data the SIA parameter $D$ can be straightforwardly estimated as $D = \lambda (g_{\parallel} - g_{\perp})/2$ \cite{AbragamBleaney2012}. With the known Ni$^{2+}$  spin-orbit coupling constant \cite{Ballhausen1962} and the above $g$-factors one gets $D \approx 0.7$\,meV. Its positive sign implies the easy-plane anisotropy of the Ni spins and its magnitude is in a quantitative agreement with the theoretical prediction \cite{Dioguardi20}.     

In contrast to an almost isotropic and gapless ESR behavior at $T \gg T_{\rm N}$, the AFM resonance in the ordered state of Ni$_2$P$_2$S$_6$ is strongly anisotopic and gapped. 
The low-energy AFM resonance branch was identified in Ref.~\cite{Mehlawat2022} [Fig.~\ref{fig:ESR_Ni2P2S6}(b)]. It shows a strong up-frequency shift from the paramagnetic position following the $\nu(H_{\rm res})$ dependence typical for an easy-plane antiferromagnet and reveals a zero-field excitation gap $\Delta = 260$\,GHz (1.07\,meV). This gapped spin excitation could not be detected by inelastic neutron scattering (INS) \cite{Lancon2018} but becomes observable due to an excellent energy resolution of the multi-frequency high-field ESR that has been appreciated in the later INS work in Ref.~\cite{Wildes2022}.

\subsubsection{Mn$_2$P$_2$S$_6$}
\label{subsubsec:MnPS}

A counterexample of the easy-plane anisotropic Ni$_2$P$_2$S$_6$ is its sister compound Mn$_2$P$_2$S$_6$ which is described as a Heisenberg antiferromagnet with an easy-axis anisotropy \cite{Joy1992,Wildes2007,Shemerliuk2021,Lu2022}. 
	\begin{figure}[h]
	\centering
	\includegraphics[width=0.6\linewidth]{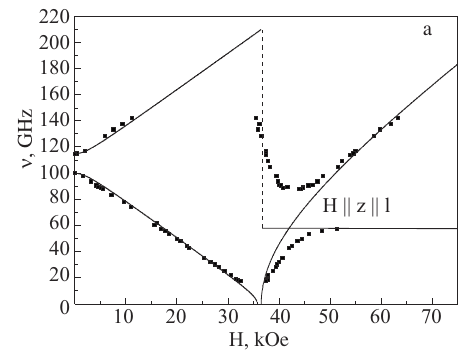}
	\caption{Frequency {\it versus} magnetic field dependence of the AFM resonance modes in Mn$_2$P$_2$S$_6$ for the $\mathbf{H}\parallel \mathbf{c^\ast}$ field geometry from Ref.~\cite{Kobets2009}. There are two types of spin excitations with different gap values at zero magnetic field $\Delta_1 = 102$\,GHz and $\Delta_2 = 116$\,GHz. Solid curves correspond to the theoretical description of the resonance branches of a rhombic biaxial two-sublattice antiferromagnet. The two branches collapse at the spin-flop field $H_{\rm sf} = 36.5$\,kOe. Above $H_{\rm sf}$ the observed branches deviate form the theoretical description which has been ascribed to their interaction resulting in the coupled vibrations \cite{Kobets2009}. (Reprinted from M. I. Kobets {\it et al.}, Low Temp. Phys. {\bf 35} (12), 930 (2009) \cite{Kobets2009} with the permission of AIP Publishing.)} 
	\label{fig:Kobets2009}
\end{figure}
Its AFM ordering temperature amounts to $T_{\rm N} = 77$\,K which is twice smaller than that of  Ni$_2$P$_2$S$_6$ suggesting  a significantly weaker coupling between the Mn spins in  Mn$_2$P$_2$S$_6$ and a smaller magnetocrystalline anisotropy that stabilizes a 3D order in van der Waals magnets.

First multi-frequency ESR experiments on single crystals of Mn$_2$P$_2$S$_6$ were reported by \citet{Okuda1986}. From measurements of the AFM resonance at $T = 4.2$\,K they concluded that Mn$_2$P$_2$S$_6$ is a uniaxial 
antiferromagnet with the zero field spin excitation gap $\Delta \approx 106$\,GHz. In the paramagnetic state, in the temperature range 110\,K\ $< T < 293$\,K the $(3\cos^2\theta - 1)^2$ type of the angular dependence of the ESR linewidth was observed an ascribed to the 2D-spin diffusion effect \cite{Benner1990}. Later, \citet{Kobets2009} reported a more detailed study of the AFM resonance at $T = 4.2$\,K that revealed two, ascending and descending  AFM modes with different zero-field gaps amounting to $\Delta_1 = 102$\,GHz a and $\Delta_2 = 116$\,GHz, respectively. These modes collapse at the spin-flop transition field $H_{\rm sf} = 36.5$\,kOe and evolve above $H_{\rm sf}$ as interacting modes causing coupled vibrations \cite{Kobets2009} (Fig.~\ref{fig:Kobets2009}). Observation of the two gaps evidences that Mn$_2$P$_2$S$_6$ is in fact a biaxial antiferromagnet and not a uniaxial one as conjectured in Ref.~\cite{Okuda1986}. 

	\begin{figure}[b]
	\centering
	\includegraphics[width=0.8\linewidth]{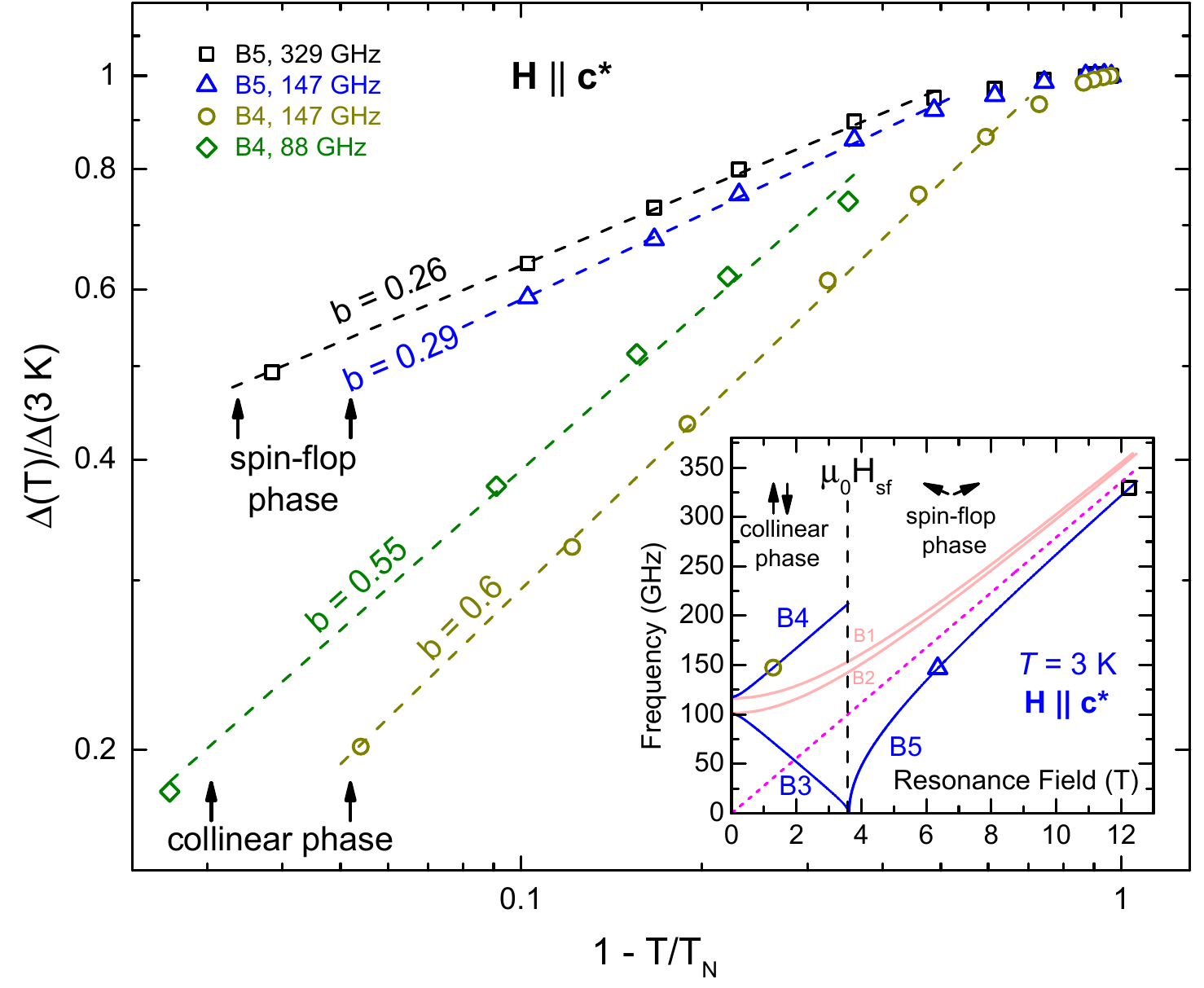}
	\caption{Main panel: Temperature dependence of the normalized energy gap $\Delta(T)/\Delta(3\,{\rm K}) = [1-(T/T_{\rm N})]^b$ for Mn$_2$P$_2$S$_6$ at different field regimes rectified from the respective dependences of the AFM resonance modes B3--B5 whose field dependence at $T = 3$\,K is schematically shown in the inset. There, the vertical dashed line separates the spin collinear phase at $H < H_{\rm sf}$ and the spin flop phase at $H > H_{\rm sf}$.
		(Reprinted with permission from J.~J.~Abraham {\it et al.}, Phys. Rev. B. {\bf 107} (16), 165141 (2023) \cite{Abraham2023}. Copyright (2023) by the American Physical Society.)} 
	\label{fig:Abraham2009}
\end{figure}

The frequency {\it versus} field diagram of the AFM resonance modes was further refined in Ref.~\cite{Abraham2023} by extending the frequency and temperature range of the study. Besides confirming the biaxial nature of the AFM spin-lattice in  Mn$_2$P$_2$S$_6$, the temperature evolution of the AFM modes was studied in some detail. Of a particular interest was the analysis of the temperature behavior of the spin excitation gap $\Delta(T)$ at $T < T_{\rm N}$ and at different magnetic fields. It can be well fitted to the power law $\Delta(T) \propto [1-(T/T_{\rm N})]^b$ (Fig.~\ref{fig:Abraham2009}). The exponent $b$ changes drastically from its value  0.55 - 0.6 in the spin collinear phase at $H<H_{\rm sf}$ to 0.26 - 0.29 in the spin-flop phase above $H_{\rm sf}$. In a quasi-2D antiferromagnet this excitation gap usually scales with the sublattice magnetization $M_{\rm sl}$ which is the AFM order parameter. Thus, $b$ can be considered as a critical exponent $\beta$ of $M_{\rm sl}$. The value of $b$ in the spin-collinear phase is close to the mean-field exponent $\beta = 0.5$, whereas in the spin-flop phase it becomes close to the critical exponent $\beta = 0.231$ in the 2D $XY$ spin model \cite{Bramwell1993}. Hence, the spin lattice in  Mn$_2$P$_2$S$_6$ demonstrates an interesting field-driven dimensional crossover of the spin excitations in the AFM-ordered state that      
boosts the effective 2D $XY$ anisotropy at $H>H_{\rm sf}$.

Signatures of the 2D spin dynamics in Mn$_2$P$_2$S$_6$ above $T_{\rm N}$ in terms of the specific angular dependence of the linewidth  $\Delta H \propto (3\cos^2\theta - 1)^2$ initially observed in Ref.~\cite{Okuda1986} at high frequencies were found also in ESR experiments at a "low" X-band frequency in Refs.~\cite{Chaudhuri2022,Senyk2023}. In particular, signatures of three different spin dynamic regimes in the paramagnetic state with distinct temperature and angular dependences of the linewidth alternating from room temperature down to $T_{\rm N}$ were observed in Ref.~\cite{Senyk2023}: prominent spin correlations of the 2D character at $T > \sim 150$\,K, a 2D\,$\rightarrow$\,3D crossover  spin dynamic regime at $\sim $ 150\,K~$> T > \sim  100$\,K followed by predominantly 3D spin correlations by approaching the long-range AFM order at $T_{\rm N} = 77$\,K. It should be noted that in the early work in Ref.~\cite{Joy1993} no signatures of the 2D spin correlations in the angular dependence of $\Delta H$ were found possibly due to a much broader width of the ESR signal of the studied crystals as compared to the narrow resonances observed in Ref.~\cite{Senyk2023}. 
%
%
%

\subsubsection{MnNiP$_2$S$_6$}
\label{subsubsec:MnNiPS}
 
Mn$_2$P$_2$S$_6$ and  Ni$_2$P$_2$S$_6$ are distinct in their anisotropic magnetic behavior featuring easy-axis and easy-plane collinear AFM order, respectively. Evolution of the magnetic properties in the series  (Mn$_{1-x}$Ni$_x$)$_2$P$_2$S$_6$  was studied by static magnetometry in Ref.~\cite{Shemerliuk2021}. The minimum ordering temperature of $T_{\rm N} \sim 57$\,K and the seemingly vanishing magnetic anisotropy was found for the compound MnNiP$_2$S$_6$. ESR spectroscopic insights into the spin excitations in the magnetically ordered ground state of this compound and spin-spin correlations at high temperatures were obtained in Refs.~\cite{Senyk2023,Abraham2023}. In contrast to Mn$_2$P$_2$S$_6$, the angular dependence of the ESR linewidth showed the $\cos^2\theta + 1 $ behavior typical for 3D spin systems possibly due to an enhanced inter-plane coupling promoted by the Ni substitution. The critical
broadening and the shift of the ESR line was observed already starting at high temperature $T \sim 200$\,K~$\gg T_{\rm N}$. It was attributed to a random distribution of Mn$^{2+}$
and Ni$^{2+}$ ions  at the $M$ site of the crystal structure of $M_2$P$_2$S$_6$ ($M_2$ = MnNi) giving rise to a competition between different types of order with contrasting magnetic anisotropies (easy-axis {\it versus} easy-plane). 

In the ordered state at $T\ll T_{\rm N}$ the measured AFM resonance branches for the external field applied normal and parallel to the $ab$-plane of the MnNiP$_2$S$_6$ crystal follow the resonance condition for a hard direction of an antiferromagnet with two different values of the zero-field excitation gap $\Delta_1 = 115\pm 9$\,GHz and $\Delta_2 = 215\pm 1$\,GHz, respectively. This observation suggests that neither of the two field directions are energetically favorable for the ordered spins implying that the magnetic structure of MnNiP$_2$S$_6$ is more complex than the collinear two-sublattice AFM model applicable to the Mn and Ni counterparts. The above gap values are intermediate between those found for Mn$_2$P$_2$S$_6$ and  Ni$_2$P$_2$S$_6$ suggesting that substitution of Mn by Ni gradually enhances magnetic anisotropy which is responsible for the opening of a gap in the spin excitation spectrum.

\subsubsection{Fe$_2$P$_2$S$_6$}
\label{subsubsec:FePS}

Fe$_2$P$_2$S$_6$ is an antiferromagnet ordering at $T_{\rm N} \approx 120$\,K in the zig-zag-type structure with the Fe spins pointing out of the $ab$-crystal plane \cite{Rule2007}. The Fe$^{2+}$ ion with 6 electrons on the 3$d$ shell occurs in the  high-spin state with $S = 2$, which is the largest spin value in the $M_2$P$_2$S$_6$ family. Moreover, Fe$_2$P$_2$S$_6$ features the largest magnetic anisotropy of the easy-axis type enabling one to classify it as an Ising antiferromagnet. 
 \begin{figure}[b!]
	\centering
	\includegraphics[width=0.8\linewidth]{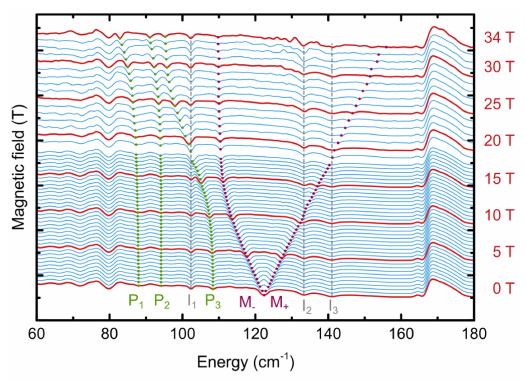}
	\includegraphics[width=0.4\linewidth]{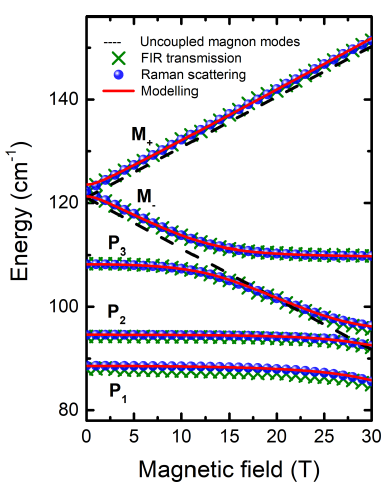}
	\caption{Top: Far-infrared transmission
		spectra of Fe$_2$P$_2$S$_6$  for selected values of the magnetic field applied perpendicular to the layer planes. $M_+$ and $M_-$ are magnetic excitations, $P_i$ are the phonon modes, and $I_i$ are other absorption resonances. Bottom: Magnetic field dependence of the energy positions of hybrid magnon-phonon modes. Crosses and circles correspond to the far-infrared transmission signals and Raman scattering peaks, respectively. Lines are the results of the modeling. (Reprinted with permission from D.~Vaclavkova {\it et al.}, Phys. Rev. B. {\bf 104} (13), 134437 (2021) \cite{Vaclavkova2021}. Copyright (2021) by the American Physical Society.)} 
	\label{fig:Vaclavkova2021}
\end{figure}

The strong magnetic anisotropy of this material implies that observation of ESR and AFM resonance requires excitation frequencies  beyond the limits of "conventional" high-frequency ESR techniques usually confined to $\nu \lesssim 1$\,THz. Indeed, the AFM resonance excitations in Fe$_2$P$_2$S$_6$ were probed with far-infrared (FIR) spectroscopy. There, the distinction between the phonon modes active in this spectral region from magnetic modes were achieved by measuring the infrared spectra in variable magnetic fields, since only the latter type of excitations  can be field sensitive.

The zero-field AFM excitation gap arising due to the Ising-like magnetic anisotropy is large and amounts according to Ref.~\cite{Vaclavkova2021} to 122\,cm$^{-1}$ (3.66\,THz). The FIR map of the observed modes is shown in Fig.~\ref{fig:Vaclavkova2021}. There, the magnetic excitations $M_\pm$ are field dependent whereas the energy of the phonon modes $P_{1,2,3}$ is constant in the moderate magnetic fields. Interestingly, at $\mu_0H > 12$\,T the magnetic mode $M_-$ demonstrates the so-called avoided crossing, first with $P_1$ and then in series with $P_2$ and $P_3$ modes. This effect is a signature of the formation of the hybrid magnon-phonon modes which suggests pronounced magneto-elastic coupling in Fe$_2$P$_2$S$_6$. The magnon-phonon hybridization in Fe$_2$P$_2$S$_6$ was further addressed in more detail in Ref.~\cite{Cui2023} where the zero-field hybridization gap between the otherwise degenerated magnon modes was observed and signatures of magnon-induced chiral phonons were identified. Interestingly, besides the one-magnon excitations giving rise to the $M_\pm$ modes Ref.~\cite{Wyzula2022} reports an observation of an exotic multipolar magnon that corresponds to the full reversal of the spin $S = 2$ of the Fe$^{2+}$ ion. Its excitation energy in zero field amounts to 57.5\,meV (13.9\,THz) and its Zeeman splitting is 4 times larger than that of the  one-magnon $M_\pm$ modes.

\subsubsection{CuCrP$_2$S$_6$}
\label{subsubsec:CuCrPS}

The quartenary compound CuCrP$_2$S$_6$ is structurally distinct from other members of the $M_2$P$_2$S$_6$ family in that nonmagnetic Cu$^{1+}$ and magnetic Cr$^{3+}$ ions are alternatively arranged on the common honeycomb plane and Cu$^{1+}$ ions are displaced up and down the plane resulting in the local electric dipole moments. This material undergoes an antiferroelectric phase transition at 140\,K \cite{Susner2020,Cho2022} and an AFM phase transition at $T_{\rm N} = 30$\,K \cite{Colombet1982,Selter2023}. Thus, CuCrP$_2$S$_6$ represents an interesting realization of interpenetrating Cu-based antiferroelectric and Cr-based antiferromagnetic lattices.  According to the static magnetic data CuCrP$_2$S$_6$ is an easy-plane antiferromagnet with an additional in-plane easy-axis along the $a$-direction \cite{Wang2023,Selter2021}. So far, little is known about ESR and AFM resonance properties of this compound. A room temperature ESR signal at the X-band frequency was measured in Ref.~\cite{Colombet1982} and attributed to the Cr$^{3+}$ ions. Recently, \citet{Wang2023} reported  mutifrequency ESR experiments in a range $\nu = 3 - 13$\,GHz. At $T > T_{\rm N}$ a single paramagnetic $\nu(H)$ resonance branch was observed, whereas at $T < T_{\rm N}$ two AFM branches were found for the $\mathbf{H} \parallel \mathbf{a}$ field geometry demonstrating a typical resonance response of an antiferromagnet with the magnetic field applied along the easy-axis. The first branch has a zero-field excitation gap of $\sim 14$\,GHz and softens towards the spin-flop field of $\mu_0H_{\rm sp}\sim 0.4$\,T. The second branch ascends out of $H_{\rm sp}$ towards stronger magnetic fields and reaches the maximum frequency of 13\,GHz at 0.6\,T.

\subsection{CrSBr}\label{subsec:CrSBr}

In CrSBr the magnetic Cr$^{3+}$ ($S = 3/2$) ions are coordinated by S and Br ligands forming the distorted CrS$_2$Br$_2$ octahedra. The interconnected octahedra build up a 2D spin-3/2 square lattice in the $ab$-crystal plane. The planes are stacked along the $c$-axis. 
%
%
The intralayer magnetic exchange is ferromagnetic while the interlayer coupling is antiferromagnetic yielding a long-range AFM order at $T_{\rm N} = 132$\,K \cite{Telford2020}. The ordered spin lattice exhibits a triaxial anisotropy with the easy $b$-axis, an intermediate $a$-axis, and the hard $c$-axis \cite{Telford2020}.     

\begin{figure}
	\centering
	\includegraphics[width=\linewidth]{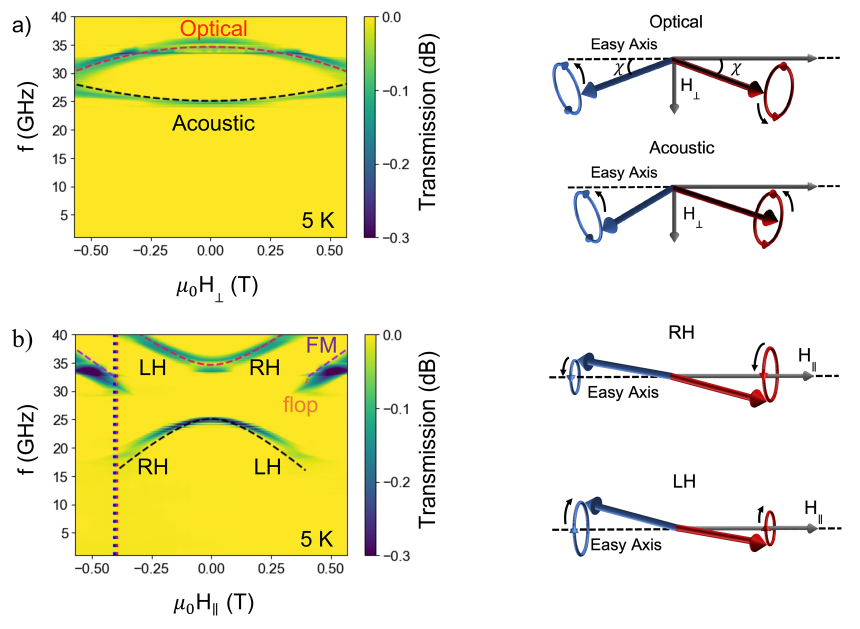}
		\caption{AFM resonance branches of CrSBr at $T = 5$\,K $\ll T_{\rm N}$ for the in-plane orientation of the applied magnetic field normal to the easy $b$-axis ($H_\perp$) (a) and parallel to it ($H_\parallel$) (b). The modes of the oscillating sublattice magnetization vectors -- optical, acoustic, right-hand (RH), and left-hand (LH) -- are illustrated on the right.
			(Reprinted with permission from T.~M.~J.~Cham {\it et al.}, Nano Lett. {\bf 22} (16), 6716 (2022) \cite{Cham2022}. Copyright (2022) by the American Chemical Society.)} 
	\label{fig:Cham2022}
\end{figure}

The specific type of anisotropy in CrSBr was reflected in the frequency \textit{versus} magnetic field dependence of the AFM modes investigated in detail in Ref.~\cite{Cham2022}. For the field applied in the $ab$-plane along the intermediate $a$-axis, i.e., normal to the easy $b$-axis ($H_\perp$) two gapped modes were observed [Fig.~\ref{fig:Cham2022}(a)]. One mode is the acoustic mode corresponding  to the in-phase oscillations of the sublattice magnetization vectors $\mathbf{M_1}$ and $\mathbf{M_2}$ of the two-sublattice collinear antiferromagnet tilted away from the easy axis by the applied field.  The other mode is the optical mode corresponding to the out-of-phase oscillations of these vectors (see the sketches in Fig.~\ref{fig:Cham2022}). Turning the field direction parallel to the easy $a$-axis ($H_\parallel$) changes the behavior of the AFM modes [Fig.~\ref{fig:Cham2022}(b)]. At small fields $\mathbf{M_1}$ and $\mathbf{M_2}$ are almost aligned along the easy axis and can oscillate in the right-hand (RH) and left-hand (LH) manner (see the sketches in Fig.~\ref{fig:Cham2022}). The presence of biaxial anisotropy in the $ab$-plane causes the avoiding crossing of the modes at zero field where they appear as symmetrical and antisymmetrical combination of RH and LH oscillations. Increasing the field strength up to $H_\parallel = \pm 0.4$\,T causes discontinuity of the $\nu(H_\parallel)$ dependence due to the flop of $\mathbf{M_1}$ and $\mathbf{M_2}$. Above this field only one mode is present corresponding to the joint precession of the two vectors to be eventually aligned along the field direction in the forced ferromagnetic (FM) state.

The observed AFM resonance branches were successfully modeled with the two coupled Landau-Lifshitz equations yielding the interlayer exchange field $\mu_0H_{\rm e} = 0.395$\,T, and the hard and
intermediate-axis anisotropy parameters $\mu_0H_{\rm c} = 1.3$\,T and $\mu_0H_{\rm a} = 0.383$\,T, respectively. The authors of Ref.~\cite{Cham2022} notice that $H_{\rm e}$ in CrSBr is significantly larger than that, e.g., in CrCl$_3$ ($H_{\rm e} = 0.1$\,T, see Sect.~\ref{subsec:CrCl3})  which may explain a higher AFM ordering temperature of CrSBr.

The out-of-plane AFM modes in CrSBr for $\mathbf{H}\parallel\mathbf{c}$ were probed in Ref.~\cite{Bae2022}. As expected for the hardest anisotropy axis, both acoustic and optical modes soften towards the saturation field $\mu_0H_{\rm sat}\approx 1.7$\,T where the sublattice magnetization vectors $\mathbf{M_1}$ and $\mathbf{M_2}$ are fully polarized in the field direction in the forced FM state. Furthermore, since CrSBr is a direct-gap semiconductor \cite{Telford2020}, it could be shown in Ref.~\cite{Bae2022} that the AFM spin waves may coherently modulate the electronic structure resulting in the magnon-exciton coupling. By means of the time-resolved pump-probe spectroscopy it was demonstrated that the energy of the excited electron-hole pair (exciton) is modulated at the frequencies of the  acoustic and optical AFM modes due to a spin-dependent interlayer electron-exchange interaction.       

AFM resonance excitations in CrSBr were further quantified in Ref.~\cite{Cho2023} where the field dependence of the two modes was measured in the three principle crystallographic directions. The authors have developed the microscopic model based on a general spin Hamiltonian compatible with the orthorhombic symmetry of CrSBr to describe the experimental data that enabled a precise determination of the AFM interlayer exchange $J_\perp = 0.069$\,K and the single-ion anisotropy parameters $D = 0.396$\,K and $E = 0.207$\,K. 

The ESR response in the paramagnetic state of CrSBr was reported in Ref.~\cite{Moro2022}. Signatures of the 2D FM spin correlations at elevated temperatures were observed in the specific angular dependence of the linewidth $\Delta H \propto (3\cos^2\theta - 1)^2$ \cite{Benner1990} where $\theta$ is the angle between the $c$-axis and the applied magnetic field. From the analysis of the angular and temperature dependences of the ESR parameters it was concluded that CrSBr can be described as a 2D magnet with the XY anisotropy, and as such it might provide a model system for the observation
of topological phase transitions and of the formation of bound vortex and antivortex pairs.

\subsection{CrPS$_4$}\label{subsec:CrPS4}

Similar to the case of  CrSBr, the Cr$^{3+}$ ions in the van der Waals compound CrPS$_4$ form a ferromagnetic 2D square spin lattice in the $ab$-crystal plane whereas the planes are weakly AFM coupled. However, in contrast to CrSBr, the Cr spins in CrPS$_4$ order below $T_{\rm N} = 34$\,K normal and not parallel to the $ab$-plane \cite{Calder2020} so that this material can be classified in first approximation as an easy-axis antiferromagnet. The weak interlayer coupling enabled \citet{Li2023} to probe AFM excitations in the ordered state of CrPS$_4$ at frequencies  up to 25\,GHz were both acoustic (in-phase) and optical (out-of-phase) modes of precession of 
the sublattice magnetization vectors $\mathbf{M_1}$ and $\mathbf{M_2}$
could be observed. It turns out that the excitation spectra for the  in-plane magnetic  field applied along $a$- and $b$-axes are different suggesting that the magnetic anisotropy of CrPS$_4$ is not uniaxial but orthorombic. Interestingly, application of the magnetic field at intermediate orientations between these two crystal axes gives rise to the hybridization of the two modes. While for $\mathbf{H}\parallel\mathbf{b}$ these modes are not entangled and cross each other, at intermediate angles hybridization yields an anticrossing gap amounting to several GHz. For $\mathbf{H}$ applied normal to the $ab$-plane  another interesting effect, a change of the sense of precession of $\mathbf{M_1}$ and $\mathbf{M_2}$ referred to as chirality switching, was observed at a critical field given by the difference of the magnetic anisotropy fields in the $a$ and $b$ directions suggesting an intimate relation of this effect to the orthorombic magnetic anisotropy in  CrPS$_4$.

\section{Ferromagnetic van der Waals compounds}\label{sec:FMvdW}

Ferromagnetic semiconducting or metallic van der Waals compounds appear to be particularly attractive for applications in magneto-electronic devices where they can be used for the generation of spin polarized currents, as an element of field-effect transistors, spin filters, etc. An important prerequisite for that is stabilization of ferromagnetic order in these compounds in the limit of a  few or single layers which can be maintained if a significantly strong easy-axis magnetic anisotropy  is present in these materials. Quantitative information on magnetic anistropy and magnetization dynamics can be obtained from measurements of ferromagnetic resonance. Since in this kind of materials magnetic layers are only weakly coupled,
%
%
reliable predictions on the magnetic behavior in the two-dimensional limit can be made from FMR measurements on bulk crystals, whereas recent progress in the detection techniques makes it possible to study FMR directly on thin flakes cleaved from a bulk crystal. 

An overview of the currently available FMR techniques and their applications to various ferromagnetic van der Waals compounds has been recently published in Ref.~\cite{Tang2023}. Here, along with mentioning selected FMR studies reviewed in Ref.~\cite{Tang2023} we will discuss also  works which were not in the scope of that review, in particular those addressing besides FMR also ESR in the paramagnetic state.

\subsection{Cr$X_3$ ($X$ = Cl, Br, I)}\label{subsec:CrX3}

In chromium-based trihalides Cr$X_3$ ($X$ = Cl, Br, I) the six-fold octahedrally coordinated Cr$^{3+}$ ions with $S = 3/2$ and $L = 0$ form the edge-sharing network in the $ab$ crystal plane resulting in a 2D magnetic honeycomb lattice. The planes are stacked along the $c$-axis and are held together via weak van der Waals forces (Fig.~\ref{fig:structure_CrX3}). The Cr spins are ferromagnetically coupled in the plane. The coupling between the planes is ferromagnetic for $X$ = Br and I and is antiferromagnetic for $X$ = Cl. A 2D in-plane FM order in CrCl$_3$ takes place at $T_{\rm c}\approx 17$\,K followed by the a inter-plane AFM order at $T_{\rm N} \approx 15$\,K \cite{Kuhlow1982}. In contrast, both CrBr$_3$ and CrI$_3$ undergo a transition to a 3D FM order at $T_{\rm c} = 37$\,K \cite{Tsubokawa1960} and 68\,K \cite{Dillon1965}, respectively. ESR spectroscopy of the antiferromagnet CrCl$_3$ was overviewed in Sect.~\ref{subsec:CrCl3}. In the following a summary of the magnetic resonance studies of ferromagnetic members of the Cr$X_3$ family will be presented.

Probably the first FMR experiment on single crystals of CrBr$_3$ was reported in 1962 by \citet{Dillon1962}. From the temperature dependence of the resonance field measured at a frequency of about 20\,GHz the temperature dependence of the easy-axis uniaxial magnetocrystalline anisotropy was determined amounting to $K = 9.4\cdot 10^5$\,erg/cm$^3$ at the minimum temperature $T = 1.5$\,K. Recent broad-band FMR study in Ref.~\cite{Shen2021} covering the frequency range $1-40$\,GHz arrived at the similar result. There, the dispersion in small magnetic fields of the in-plane FMR mode was successfully fitted in the frame of the multi-domain FMR theory.  

Critical spin dynamics by approaching the ordering temperature $T_{\rm c}$ from above was analyzed in Refs.~\cite{Alyoshin1997,Saiz2021,Clemente2023} from the temperature dependent critical broadening of the ESR linewidth $\Delta H_{\rm crit} \propto [(T - T_{\rm c})/T_{\rm c}]^{-p}$ at 0.36\,GHz \cite{Alyoshin1997}, 9.4\,GHz \cite{Clemente2023} and 240\,GHz \cite{Saiz2021}.  Interestingly, despite a large difference in the excitation frequencies and corresponding external magnetic fields the critical exponent $p$ converges to the range $p \sim 0.5 - 0.9$ in these works.  In purely 2D magnets the critical exponent $p$ is expected to be mostly larger than 1 and to sensitively depend on the type of the exchange interaction, FM or AFM, and should be distinct for Heisenberg, Ising or $XY$ magnets \cite{Benner1990}.  However, in real systems the finite interlayer interaction hampers the 2D critical behavior by this reducing $p$ in most experimental cases (FM or AFM) to $p < 1$ \cite{Benner1990}, as it apparently the case for CrBr$_3$, too.

In fact, the two-dimensionality of the spin correlations in CrBr$_3$ manifests in the $(3\cos^2\theta - 1)^2$ type of the angular dependence of the ESR linewidth as observed in Refs.~\cite{Saiz2021,Clemente2023}. This kind of dependence is due to the dominating secular part of the dipole-dipole interaction which for the case of 2D FM correlations has the angular form $(3\cos^2\theta - 1)$ \cite{Benner1990}. In 3D magnets the angular dependence of the linewidth is typically of the form $\Delta H \propto (\cos^2\theta + 1)$ \cite{Benner1990}.

In CrI$_3$, FMR was first probed by \citet{Dillon1965} in 1965 at three excitation frequencies of 86, 91 and 99\,GHz and at a temperature of 1.5\,K. The analysis of the data yielded the easy-axis uniaxial magnetocrystalline anisotropy constant $K = 3.1\cdot 10^6$\,erg/cm$^3$. Its value appears three times larger as the one for the sister compound, CrBr$_3$. Since magnetic anisotropy is an important contributor to the stabilization of magnetic order in quasi-2D systems, this result is consistent with a much higher FM ordering temperature $T_{\rm c}$ in CrI$_3$ as compared to CrBr$_3$. These early experiments were later significantly extended regarding the frequency  and temperature range where the magnetic resonance excitations were probed \cite{Shen2021,Jonak2022}, providing more accurate estimates of the relevant magnetic parameters. While the work in Ref.~\cite{Shen2021} focuses on the low-frequency domain 1--40\,GHz, Ref.~\cite{Jonak2022} reports FMR results in a frequency range 30--330\,GHz in magnetic fields up to 15\,T. Remarkably, in the latter work measurements at $T > T_{\rm c}$ revealed the persistence of a zero field excitation gap and quasi-static anisotropic internal fields up to $T\approx 1.3\,T_{\rm c}$ thereby suggesting that a significant easy-axis anisotropy maintains 2D short-range magnetic order in CrI$_3$ above the bulk FM ordering temperature.

Detailed insights into the structure of the magnetic anisotropy of CrI$_3$ were obtained in Ref.~\cite{Lee2020} by studying the angular dependence of the FMR signals at $\nu = 120$ and 240\,GHz  and at $T = 5$\,K. The symmetry-based theoretical analysis of the data revealed a very strong intra-plane Ising-like Kitaev interaction \cite{Kitaev2006} of $K \sim 60$\,K (-5.2\,meV), 25 times stronger than the isotropic Heisenberg exchange $J \sim - 2.3$\,K (-0.2\,meV). It is argued that this particularly strong anisotropy explains the robust FM order in a monolayer of CrI$_3$ at $T_{\rm c}^{\rm ML} = 34$\,K \cite{Huang2017}. Indeed, in Ref.~\cite{Zhang2020} an optical pump/magneto-optical Kerr probe technique was used to detect sub-THz magnetic resonance modes in CrI$_3$ bilayers under an in-plane magnetic field. These modes were successfully modeled with Landau-Lifshitz-Gilbert equations yielding the parameters, such as the zero-field excitation gap, anisotropy field, and the saturation field, rather similar to the results of the FMR measurements on bulk single crystals.      

Finally, the photosensitivity of the ESR signal of CrI$_3$ was observed by \citet{Singamaneni2020} manifesting in the change of its intensity, $g$-factor, and the linewidth under application of visible light. The authors observed the similar effect also for the antiferromagnetic CrCl$_3$ and ascribe both observations to the photo-induced electron transitions between the valence band and localized Cr$^{2+}$ ($3d^4,\, S = 2$) levels (see Sect.~\ref{subsec:CrCl3}).

\subsection{Cr$_5$Te$_8$}\label{subsec:Cr5Te8}

Cr$_5$Te$_8$ is a half-metallic van der Waals ferromagnet with a high ordering temperature $T_{\rm C}$ in the range 245--255\,K depending on the degree of the Cr deficiency \cite{Lukoschus2004,Chen2021b}. The Cr atoms are octahedrally coordinated by the Te anions and within the $ab$-layers the CrTe$_6$ octahedra share common edges \cite{Lukoschus2004}. At $210\,{\rm K} < T < T_{\rm C}$ the magnetic easy axis lies in the $ab$-plane and reorients normal to the plane below 210\,K \cite{Chen2021b}. Details of this reorientation transition were addressed in the FMR study performed in Ref.~\cite{Chen2021b} at an X-band frequency of 9.45\,GHz. Angular dependent measurements of the FMR signal were carried out in the temperature range 180--260\,K and analyzed in the frame of the phenomenological FMR  theory \cite{Smit1955,Skrotskii1966,Farle1998}. It was concluded that the reorientation of the effective magnetic anisotropy axis taking place in the range from $T_{\rm C}$ to 210\,K is due the temperature-driven competition between the 2-nd order intrinsic magnetocrystalline anisotropy of Cr$_5$Te$_8$ and the shape anisotropy of the platelet-like sample. 

\subsection{Cr$_2X_2$Te$_6$ ($X$ = Si, Ge)}\label{subsec:Cr2X2Te6}

	\begin{figure}
	\centering
	\includegraphics[width=0.8\linewidth]{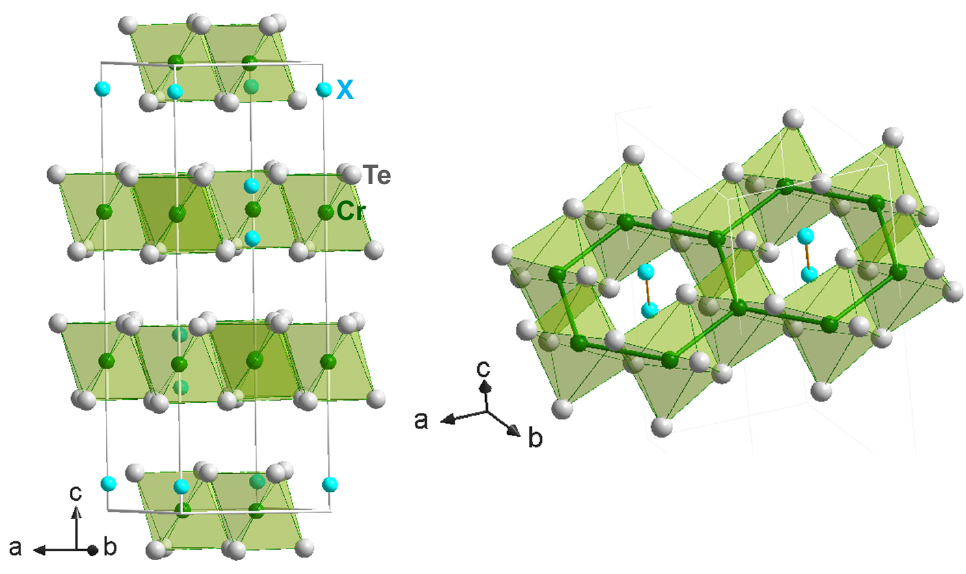}
	\caption{Layered crystal structure of the Cr$_2X_2$Te$_6$ ($X$ = Si, Ge) compounds of the rhombohedral symmetry, space group $\bar{R}3$ (left). In the individual layers CrTe$_6$ octahedra share edges and form a honeycomb lattice with the pair of $X$ atoms in the center of each honeycomb (right).} 
	\label{fig:structure_Cr2X2Te6}
\end{figure}

The main structural features of the van der Waals compounds Cr$_2X_2$Te$_6$ ($X$ = Si, Ge)  bear similarities with the $M_2$P$_2$S$_6$ family (see Sect.~\ref{subsec:MPS}). The octahedrally coordinated Cr$^{3+}\,(S = 3/2)$ cations form honeycomb layers in the $ab$-plane stacked along the $c$-axis (Fig.~\ref{fig:structure_Cr2X2Te6}). However, the magnetic properties are very different. In both Cr$_2$Si$_2$Te$_6$ and Cr$_2$Ge$_2$Te$_6$ the Cr spins are coupled ferromagnetically in the planes and between the planes and order FM out-of-plane at $T_{\rm C} = 32$\,K and 61\,K, respectively \cite{Carteaux1995a,Carteaux1995b}. In the ordered state Cr$_2$Si$_2$Te$_6$ demonstrates 2D Ising-like  magnetic behavior \cite{Carteaux1995a} while it is of the anisotropic Heisenberg-like type in Cr$_2$Ge$_2$Te$_6$ \cite{Carteaux1995b}.

The intrinsically 2D character of the spin dynamics in Cr$_2$Si$_2$Te$_6$ was characterized by a multi-frequency ESR/FMR study in the range 13--20\,GHz in Ref.~\cite{Li2022}. The authors focused on the temperature and angular dependence of the $g$-factor shift of the resonance signal both above and below $T_{\rm c}$. They found out that the $g$ shift follows the $(3\cos^2\theta - 1)$ dependence in a broad temperature range while varying the angle $\theta$ which the applied field makes with the normal to the $ab$-plane. As has been shown before by Nagata {\it et al.} (see, e.g., Ref.~\cite{Nagata1977}) and further elaborated in Ref.~\cite{Li2022} such kind of dependence arises due to an anisotropic spin-spin correlations in a 2D magnet.

Much more attention was given to magnetic resonance investigations on Cr$_2$Ge$_2$Te$_6$. This interest was stirred up by the discovery of intrinsic ferromagnetism in the few layers thin flakes of Cr$_2$Ge$_2$Te$_6$ that order below $\sim 30$\,K \cite{Gong2017}. Since according to the Mermin-Wagner theorem stabilization of magnetic order in a 2D spin system could only be possible in the presence of easy-axis anisotropy \cite{Mermin1966}, the aspects of magnetic anisotropy  in  Cr$_2$Ge$_2$Te$_6$ were extensively studied by ESR and FMR techniques. 

First efforts to quantify the magnetic anisotropy energy in bulk crystals of Cr$_2$Ge$_2$Te$_6$ were undertaken by \citet{Zhang2016} with FMR experiments in a limited frequency range $\leq 20$\,GHz. From the analysis of the in-plane FMR branch $\nu(H_{\rm ab})$ the uniaxial anisotropy constant $K_{\rm u}$ was estimated to reach $3.65\times 10^5$\,erg/cm$^3$ at the minimum temperature $T = 5$\,K. This work was followed by a comprehensive ESR and FMR investigation of the magnetic anisotropy   carried out in Ref.~\cite{Zeisner2019} over a wide frequency and temperature range. Angular and frequency dependences of the resonance modes at $T<T_{\rm C}$ were modeled using the phenomenological FMR theory \cite{Smit1955,Skrotskii1966,Farle1998}. This enabled an accurate determination of the easy-axis type of $K_{\rm u}= (4.8 \pm 0.2)\times 10^5$\,erg/cm$^3$. Density functional calculations performed in the same work could successfully reproduce this result assuming the presence of a substantial electronic correlation strength $U\sim 2$\,eV. Furthermore, multifrequency  ESR measurements in the paramagnetic state of Cr$_2$Ge$_2$Te$_6$ \cite{Zeisner2019} revealed the onset of spin-spin correlations far above $T_{\rm C}$ suggesting an intrinsic 2D character of ferromagnetism manifesting even in the bulk form of the Cr$_2$Ge$_2$Te$_6$ compound. Similar results albeit with somewhat smaller value of the estimated $K_{\rm u}$ were obtained in the FMR study at $T< T_{\rm C}$ in Ref.~\cite{Khan2019}. The observed deviation of the $g$-factor tensor elements of Cr$^{3+}$ ions to the values larger than $g = 2$ were ascribed in Ref.~\cite{Khan2019} to the transferred spin-orbit coupling of the $p$ orbitals of the heavy Te ligands hybridized with the Cr 3$d$ orbitals. The contribution of orbital magnetism to the $g$-tensor was also proposed in the FMR work in Ref.~\cite{Wang2023b} where the enhanced spin-orbit coupling was likewise  conjectured to be the reason for the observed dumping of the FMR signal.

\begin{figure}
	\centering
	\includegraphics[width=\linewidth]{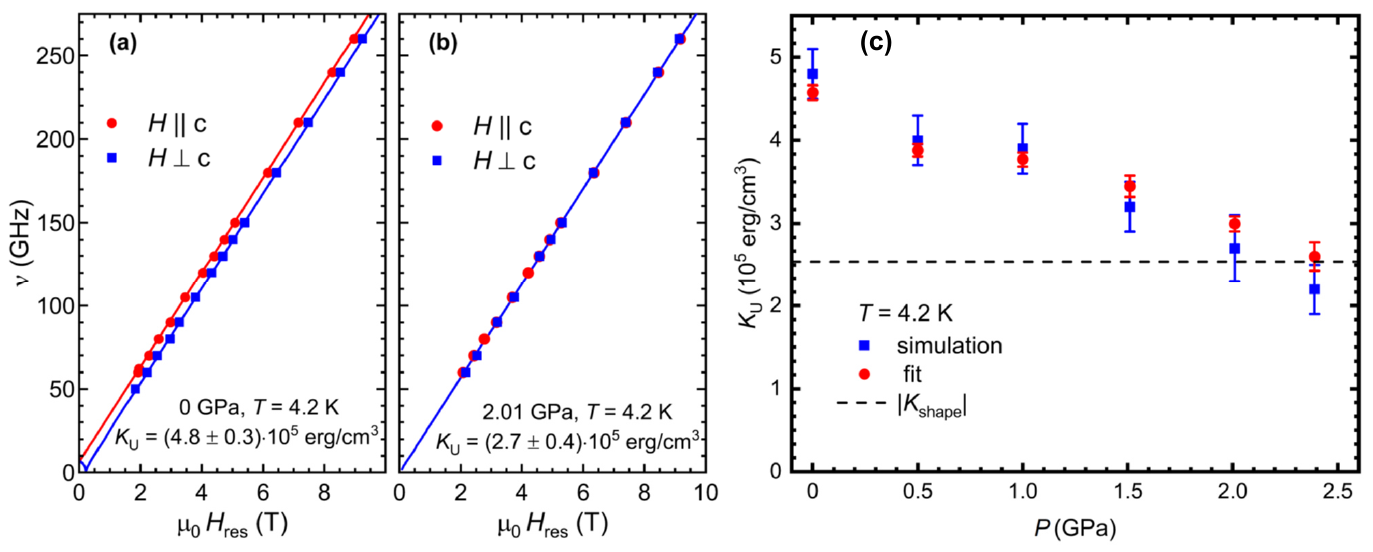}
	\caption{(a-b) FMR out-of-plane and in-plane branches $\nu(H_{\rm res})$ for Cr$_2$Ge$_2$Te$_6$ at $P = 0$\,GPa (a) and 2.01\,GPa (b). With increasing pressure the two branches come closer together due to the reduction of the magnetic anisotropy. (c) Pressure dependence of the uniaxial magnetic anisotropy constant $K_{\rm u}$ obtained from the fit of the FMR branches (red circles) and from their simulations based on the FMR theory \cite{Smit1955,Skrotskii1966,Farle1998} (blue squares). Horizontal dashed line denotes the modulus of the negative easy-plane shape anisotropy constant $K_{\rm shape}$ given by the platelet-like shape of the measured crystal. By approaching the maximum pressure  $P = 2.39$\,GPa $K_{\rm shape}$ practically compensates $K_{\rm u}$ yielding isotropic FMR response. 
	(Reprinted with permission from T.~Sakurai {\it et al.}, Phys. Rev. B. {\bf 103} (2), 024404 (2021) \cite{Sakurai2021}. Copyright (2021) by the American Physical Society.)} 
	\label{fig:pressure_Cr2Ge2Te6}
\end{figure}

The control of the magnitude  of the magnetic anisotropy energy of Cr$_2$Ge$_2$Te$_6$ by hydrostatic pressure and its monitoring by FMR spectroscopy was achieved by \citet{Sakurai2021}. The in- and out-of-plane FMR branches $\nu(H_{\rm ab})$ and $\nu(H_{\rm c})$ were measured at $T = 4.2$\,K in the frequency range 50--260\,GHz and variable applied pressures in the range $P = 0-2.39$\,GPa [Fig.~\ref{fig:pressure_Cr2Ge2Te6}(a-b)]. It has been shown that  increasing the pressure reduced the zero field excitation gap signaling a reduction of the uniaxial magnetic anisotropy constant $K_{\rm u}$ with respect to its value of $4.8\times 10^5$\,erg/cm$^3$ at the ambient pressure. Eventually, at the highest $P = 2.39$\,GPa $K_{\rm u}$ reduced almost twice to $2.6\times 10^5$\,erg/cm$^3$ [Fig.~\ref{fig:pressure_Cr2Ge2Te6}(c)]. At this pressure the studied sample showed seemingly isotropic FMR behavior with the gapless and degenerated $\nu(H_{\rm ab})$ and $\nu(H_{\rm c})$ resonance branches because the reduced $K_{\rm u}$ was fully compensated by the negative easy-plane shape anisotropy of the platelet-like sample [Fig.~\ref{fig:pressure_Cr2Ge2Te6}(c)].

Unlike the above cited works dealing with bulk single crystals of  Cr$_2$Ge$_2$Te$_6$, \citet{Zollitsch2023} succeeded to detect and analyze the FMR modes in the frequency range 12--18\,GHz in ultra-thin flakes of this material with the lateral size in the $\mu$m range and thicknesses varying from 153\,nm down to 11\,nm (15 monolayers). The key element of their setup was a superconducting resonator with a high quality factor and strong oscillating microwave field. A strong coupling regime between the spin wave modes in the sample and microwave photons in the resonator was achieved resulting in the so-called light-matter hybrid modes or polaritons. The analysis of these hybrid modes enabled important conclusions that the magnetic anisotropy constant $K_{\rm u}$ practically did not depend on the thickness of the flake and that the Gilbert dumping parameter in a 15 monolayers thin flake $\alpha\lesssim 0.021$ appeared to be comparable with conventional ferromagnetic films, such as FeNi (permalloy), altogether making Cr$_2$Ge$_2$Te$_6$ in the 2D limit a promissing candidate for spintronic applications.  

\subsection{Fe$_x$GeTe$_2$ ($x$ = 3, 4, 5)}\label{subsec:FeGeTe}

The family of Fe-rich  van der Waals compounds Fe$_x$GeTe$_2$ ($x$ = 3, 4, 5) feature metallic conductivity, sizable magnetic anisotropy and high ferromagnetic ordering temperatures $T_{\rm C}$ approaching and even exceeding room temperature. Therefore, they are considered for the use in room-temperature spintronic devices. For this reason complex magneto-transport properties of Fe$_x$GeTe$_2$ are currently in the focus of intensive research while there are only a few reports on  FMR studies on these materials.

Ref.~\cite{Ni2021} reports FMR experiments at the X-band frequency of 9.48\,GHz on a single crystal of Fe$_3$GeTe$_2$ with $T_{\rm C}\approx 204$\,K. Broad FMR absorption line of the order of $\sim 2$\,kOe was observed whose angular and temperature dependence was analyzed with the conventional FMR theory \cite{Smit1955,Skrotskii1966,Farle1998}. As a result the effective dumping parameter was estimated amounting to $\alpha_{\rm eff} \sim 0.58$. The authors conjecture that its much larger value as, e.g., for Cr$_2$Ge$_2$Te$_6$ (see Sect.~\ref{subsec:Cr2X2Te6}) could be due to the itinerant character of ferromagnetism of Fe$_3$GeTe$_2$, enhanced magnon scattering or spatial magnetization inhomogeneity across the sample. 

Broadband FMR spectroscopy on a single crystal of Fe$_5$GeTe$_2$ with $T_{\rm C} = 332$\,K was carried out in the frequency range 5--150\,GHz at various temperatures between 10 and 300\,K in Ref.~\cite{Alahmed2021}. The data were analyzed using the standard theoretical FMR approach. The authors find a significant anisotropy of the $g$-factor suggesting contribution of the orbital magnetism possibly due to the hybridization of the Fe 3$d$ orbitals with the 5$p$ orbitals of Te. Interestingly, the effective dumping constant was evaluated to be $\alpha_{\rm eff} = 0.035$ at $T = 300$\,K and $\alpha_{\rm eff} = 0.007$ at $T = 10$\,K, i.e., drastically smaller as the one reported for Fe$_3$GeTe$_2$ in  Ref.~\cite{Ni2021} and comparable with $\alpha_{\rm eff}$ for Cr$_2$Ge$_2$Te$_6$ and permalloy (see Sect.~\ref{subsec:Cr2X2Te6}).

%
%

Very recently \citet{Pal2024} reported detailed multi-frequency FMR results on a single crystal of Fe$_4$GeTe$_2$ which orders at $T_{\rm C}\approx 270$\,K. This material attracts significant attention with respect to its transport properties featuring emergent electronic phases \cite{Pal2024b} and regarding a peculiar temperature driven reorientation of the direction of the spins at $T_{\rm SR} \sim 110$\,K from in-plane at $T > T_{\rm SR}$ to out-of-plane at $T < T_{\rm SR}$ \cite{Seo2020,Wang2023c}. FMR measurements were carried out in a frequency range 75--350\,GHz, at temperatures between 300 and 3\,K for the in-plane and out-of-plane orientation of the applied magnetic field. An analysis of the temperature evolution of the FMR $\nu(H)$ branches made it possible to disentangle different contributions to the internal anisotropy field, to single out and quantify the temperature dependence of the intrinsic magnetocrystalline anisotropy $K_{\rm u}(T)$ of Fe$_4$GeTe$_2$. It turns out to be always of the easy-axis type in the entire temperature range but strongly increasing in the magnitude below $\sim 150$\,K and overcoming the (extrinsic) easy-plane shape anisotropy of the platelet-like crystal at $T_{\rm SR}$ which causes the spins to turn from the in-plane to the out-of-plane direction. Characteristic temperatures for the evolution of $K_{\rm u}(T)$ were found to be in a close correspondence with the emergence of electronic transitions observed in Ref.~\cite{Pal2024b} suggesting remarkable intertwined magnetic and electronic behaviors in Fe$_4$GeTe$_2$.

\section{Magnetic van der Waals topological insulators}\label{sec:Mag_vdW_TI}	
\begin{figure}[b]
	\centering
	\includegraphics[width=0.8\linewidth]{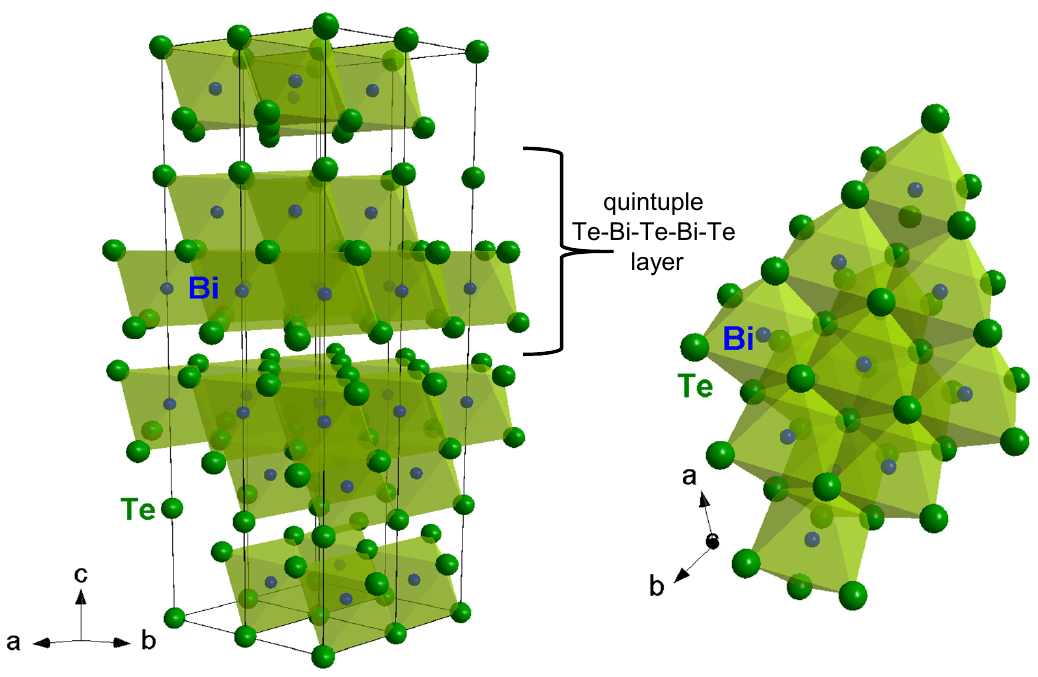}
	\caption{Crystal structure of Bi$_2$Te$_3$ in two projections. The Bi atoms are bonded to the Te atoms in the octahedral fashion and build up the quintuple layer Te--Bi--Te--Bi--Te. These layers are stacked along the $c$-axis.     } 
	\label{fig:structure_Bi2Te3}
\end{figure}

Topological insulators (TIs) are a special class of band insulators where by virtue of the strong spin-orbit coupling the order of the electronic bands in the bulk of the material is inverted (for the reviews see, e.g., Refs.\cite{Hasan2010,Hasan2011,Qi2011}). Since the spin-orbit coupling is present only in the interior, the order of the bands should be restored outside the material. Thus the bands inevitably cross at the Dirac point at the surface yielding robust, so-called topologically protected conducting electronic surface states. These surface electrons have unique properties distinct from the "trivial" two-dimensional metals. They are characterized by the gapless Dirac-cone linear energy dispersion and the spin-momentum locking which suppresses the scattering of the Dirac electrons and makes the surface of a TI highly conducting.  Due to these special properties the energy gap  at the Dirac point can only be opened by breaking the time reversal symmetry by a magnetic field \cite{Kou2015}. Such a field can be generated by doping a TI with magnetic atoms or a TI may itself possess an intrinsic regular magnetic lattice. Opening of the gap would enable new exotic phenomena such as an anomalous quantum Hall effect or topological magnetoelectric effect \cite{Qi2011,Kou2015} which bring magnetic TIs into the focus of a very active current research. 

The most well-known bulk magnetic TIs are van der Waals compounds such as magnetically doped Sb$_2$Te$_3$, Bi$_2$Te$_3$ and Bi$_2$Se$_3$, and intrinsically magnetic (MnBi$_2$Te$_4$)(Bi$_2$Te$_3$)$_{\rm n}$. Their magnetic properties were addressed in several ESR studies which will be summarized in the following.

\subsection{Magnetically doped topological insulators}\label{subsec:Mag_doped_vdW_TI}

The van der Waals compounds Bi$_2$Se$_3$ and Bi$_2$Te$_3$ termed tetradymites are isostructural and crystallize in a lattice of rhombohedral symmetry, space group $R\bar{3}m$. The building blocks of the structure are quintuple Te--Bi--Te--Bi--Te layers in the $ab$-crystal plane stacking along the $c$-axis (Fig.~\ref{fig:structure_Bi2Te3}). In fact, these materials are not really bulk insulators but rather doped semiconductors of $n$- and $s$-type for Bi$_2$Se$_3$ and Bi$_2$Te$_3$, respectively \cite{Cava2013}. Thus, besides featuring the surface conducting states \cite{Hasan2010} they are also conducting in the bulk. 

\begin{figure}
	\centering
	\includegraphics[width=0.8\linewidth]{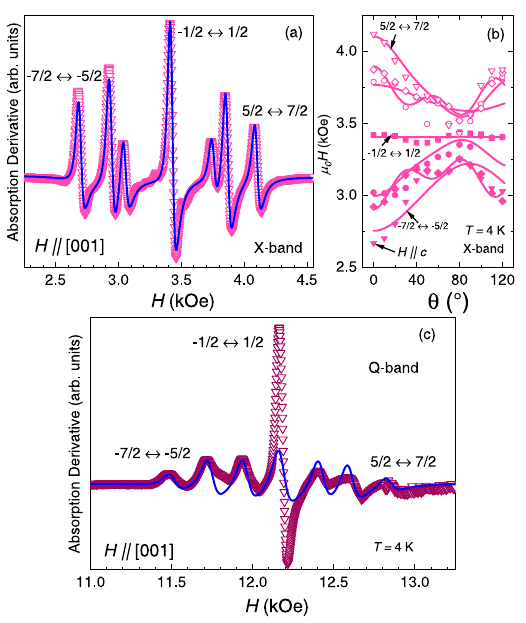}
	\caption{(a) X-band ($\nu = 9.5$\,GHz) Gd$^{3+}$ ESR spectrum for
		(Bi$_{0.994}$Gd$_{0.006}$)$_2$Se$_3$ and (b) its resonance field angle dependence at
		$T = 4\,$K with the applied magnetic field H parallel to the [001]
		direction. In panel (b) the field is rotated towards the $ab$-plane. (c) Q-band ($\nu = 34$\,GHz) Gd$^{3+}$ ESR spectrum. Solid lines are simulations based on the crystal field Hamiltonian. 
		(Reprinted with permission from J.~C.~Souza {\it et al.}, Phys. Rev. B. {\bf 106} (23), 235109 (2022) \cite{Souza2022}. Copyright (2022) by the American Physical Society.)} 
	\label{fig:Souza2022}
\end{figure}

The bulk electronic properties of Bi$_2$Se$_3$ were probed in Refs.~\cite{Garitezi2015,Souza2022} by ESR of small amount of Gd$^{3+}$ ions doped to Bi$_2$Se$_3$. The ESR spectrum (Fig.~\ref{fig:Souza2022}) features a typical fine structure due to the small splitting of the $S=5/2$ ground state multiplet of the Gd$^{3+}$ ion by the crystal field \cite{AbragamBleaney2012}. The shape of the lines is asymmetric of the Dysonian type, typical for metallic samples with dimensions larger than the penetration depth (skin depth) of the microwaves \cite{Barnes1981}.    
The width of the individual lines of the spectrum $\Delta H(T)$ depended approximately linear on temperature due to the so-called Korringa relaxation mechanism of localized spins on conduction electrons with the slope \cite{Barnes1981}:
\begin{equation}
	b = d(\Delta H)/dT \propto [D(E_{\rm F})J_{\rm ce}]^2\ .
\label{eq:KorringaSlope}
\end{equation} 
Here, $D(E_{\rm F})$ is the density of electronic states at the Fermi energy and $J_{\rm ce}$ is the constant of exchange interaction between the localized spin and the spin of a conduction electron.
Such a behavior was reasonably ascribed to the interaction of Gd spins with bulk conduction electrons whereas a significant increase of this slope below 40\,K and of the related shift of the $g$-factor $\Delta g = D(E_{\rm F})J_{\rm ce}$ \cite{Barnes1981} observed in Ref.~\cite{Garitezi2015} was speculatively attributed to the contribution of the highly conducting surface state which considering the small skin depth of the studied sample might be sizable. This effect was not discussed anymore in the later work in Ref.~\cite{Souza2022} where the authors extended the ESR experiments to the Q-band frequency of 35\,GHz. As compared to the X-band, the Q-band Gd$^{3+}$ ESR spectrum showed a strong enhancement of the central line with respect to the other components of the fine structure possibly indicating a collapse of the fine structure (Fig.~\ref{fig:Souza2022}). It was conjectured, that the conduction electrons surrounding Gd$^{3+}$ ions might be stronger localized at a smaller magnetic field corresponding to the X-band frequency and became more mobile in a higher magnetic field corresponding to the Q-band frequency and thus screen the crystal field responsible for the fine structure splitting. This was discussed  as a possible manifestation of the weak antilocalization effect \cite{Lu2014} previously observed in the magneto-transport measurements on thin layers of Bi$_2$Se$_3$ \cite{Kim2011}.

With the idea to influence the topological surface states in Bi$_2$Te$_3$  \citet{Hor2010} synthesized a series of Bi$_{\rm 2-x}$Mn$_{\rm x}$Te$_3$ single crystals with the concentration of  Mn$^{2+}$ dopants in the range $x = 0.005 - 0.09$. Static magnetometry results evidenced ferromagnetic phase transition  for samples with $x \geq 0.04$ at $T_{\rm C} \approx 12$\,K and surface sensitive angular resolved photoemission spectroscopy measured at a slightly higher temperature revealed modification of the surface states in these samples. The occurrence of FM order at such moderate concentration of magnetic dopants was rather unexpected. As a possible reason for it a long-range Ruderman-Kittel-Kasuya-Yosida (RKKY) interaction \cite{Kittel1969} between Mn spins mediated by bulk mobile charge carriers was proposed and first-principle calculations gave theoretical estimates of its strength \cite{Henk2012,Vergniory2014}. To verify theoretical  predictions detailed X-band ESR measurements on these crystals were carried out in Ref.~\cite{Zimmermann2016}. A single exchange-narrowed resonance line ascribed to Mn$^{2+}$ ions was observed for all samples and its temperature evolution was studied in the range from 80\,K down to $T_{\rm C}$. Above $\sim 25$\,K the linewidth depended approximately linear on $T$ with the Korringa slope $b$ [Eq.~(\ref{eq:KorringaSlope})] varying only slightly with $x$ around $b \sim 1$\,mT/K. At lower temperatures the signal demonstrated strong critical broadening of the 2D character due to the slowing down of the FM correlations in a quasi-2D magnet \cite{Benner1990}. Magneto-transport measurements perfomed in the same work enabled one to determine the charge-carrier density and from it to make an estimate of the density of states $D(E_{\rm F})$ and using the experimental value of $b$ to estimate, according to Eq.~(\ref{eq:KorringaSlope}), the exchange constant $J_{\rm ce} \approx 0.5-0.7$\,eV. Finally, the RKKI exchange constant $J_{\rm ij}$ between Mn spin sites $i$ and $j$ was evaluated according to \cite{Kittel1969}:     
\begin{equation}
	J_{\rm ij} =  \frac{3m^* V^2}{4\pi h^2}\cdot J_{\rm ce}^2 \cdot \frac{\sin(2k_{\rm F}r) -2k_{\rm F}r\cos(2k_{\rm F}r) }{r^4} \ .
	\label{eq:RKKY}
\end{equation}
Here, $m^*= 0.35m_{\rm e}$ is the effective mass \cite{Koehler1976}, $V$ is the unit cell volume, $k_{\rm F}$ is the Fermi vector, and $r$ is the distance between the Mn sites $i$ and $j$ which was estimated from the concentration of Mn in the samples.  $J_{\rm ij}$ appeared to be ferromagnetic with the magnitude $|J_{\rm ij}|\approx 2-3$\,meV (23--35\,K). This value agreed well with the theoretical predictions in Refs.~\cite{Henk2012,Vergniory2014} validating the long-range RKKY interaction to be responsible for the FM order of the Mn-doped Bi$_2$Te$_3$.

The development of the FM correlations by approaching $T_{\rm C}\approx 12$\,K in the Bi$_{\rm 1.95}$Mn$_{\rm 0.05}$Te$_3$ single crystal was further studied by X-Band ESR in Ref.~\cite{Talanov2017}. Measurements of the angular dependence of the Mn ESR signal revealed a practically isotropic behavior at $T > 30$\,K. By lowering the temperature towards $T_{\rm C}$ the resonance field was getting progressively anisotropic deviating from the high-temperature paramagnetic position to smaller values for the external field applied along the $c$-axis and to the higher values for the field applied in the $ab$-plane. Based on these observations the authors provided qualitative arguments for the easy-axis type of FM order in Mn-doped Bi$_2$Te$_3$ in agreement with the ESR results in Ref.~\cite{Zimmermann2016}. In addition to the main Mn ESR line the authors reported an additional weak signal in the spectrum at a low field of $\sim 1000$\,Oe for the in-plane field direction which position does not change with temperature and is not sensitive to $T_{\rm C}$. This signal was attributed to a small amount of Mn clusters forming in the host material due to the limited solubility of Mn.   

In the subsequent works in Refs.~\cite{Sakhin2018,Sakhin2019,Sakhin2021} the same group of authors  reported X-band ESR results on single crystals of the 3D topological insulator Bi$_{1.08}$Sn$_{0.02}$Sb$_{0.9}$Te$_2$S \cite{Kushwaha2016} where they observed signals attributed to intrinsic magnetic centers of different origin in the material formally composed of nonmagnetic elements. Based on the measurements of the temperature and angular dependences of the ESR spectra it was concluded that one type of the centers could be induced by the nonmagnetic structural defects around which superparamagnetic nano-sized inclusions  might arise. The other type of centers contributed to the ESR spectrum  was conjectured to be the bulk charge carriers organized in small electron and hole droplets.

Besides Mn-doped Bi$_2$Te$_3$, magnetic doping of the sister compound Bi$_2$Se$_3$ was also considered to be promising for the alteration of the band structure of the topological surface states. That moderate Mn doping can result in a long-range FM order in a thin film of Bi$_2$Se$_3$ was shown by \citet{Savchenko2018} who fabricated single-crystalline films of this material with thicknesses in the range 0.3--0.7\,$\mu$m nominally doped with 6--8\,\% of Mn. These films demonstrated an FM transition at $T_{\rm C} = 5.3$\,K and the properties of the FM ordered state were examined by X-band ($\nu = 9.4$\,GHz) ESR/FMR spectroscopy. Strong temperature and angular dependences of the FMR signal were observed and analyzed with the standard FMR theory. It was found that the films feature an easy-plane magnetic anisotropy which increases with increasing the Mn concentration and reaches the value of $K_{\rm u} = -3720$\,erg/cm$^{3}$ at the highest doping level.

\subsection{Intrinsically magnetic topological insulators}\label{subsec:Mag_intrins_vdW_TI}
\begin{figure}
	\centering
	\includegraphics[width=0.8\linewidth]{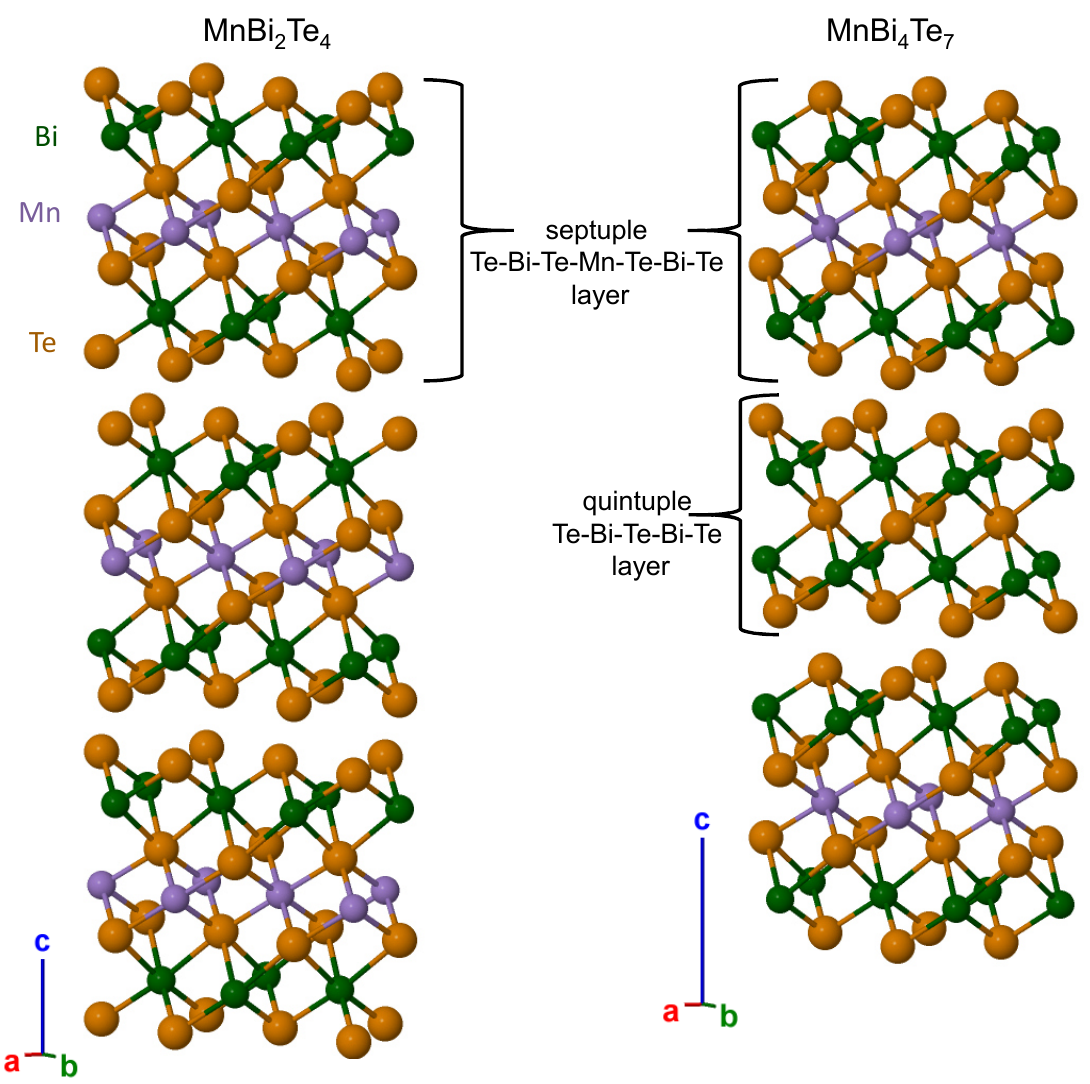}
	\caption{Crystal structure of the (MnBi$_2$Te$_4$)(Bi$_2$Te$_3$)$_{\rm n}$ family with $n = 0$ (left) and $n = 1$ (right). MnBi$_2$Te$_4$ on the left is composed of the stacked septuple Te--Bi--Te-Mn--Te--Bi--Te layers. In MnBi$_4$Te$_7$ on the right the neighboring septuple layers are separated by the quintuple Te--Bi--Te--Bi--Te layers. } 
	\label{fig:structure_Mn-Bi-Te}
\end{figure}

While magnetically doped topological insulators were known for already some time, intrinsic magnetic TIs, i.e., stoichiometric compounds comprising a regular long-range ordered magnetic lattice, were discovered only recently in the family of (MnBi$_2$Te$_4$)(Bi$_2$Te$_3$)$_{\rm n}$ van der Waals compounds \cite{Otrokov2019,Gong2019}. The crystal structures of the two mostly studied so far materials MnBi$_2$Te$_4$ and MnBi$_4$Te$_7$ are shown in Fig.~\ref{fig:structure_Mn-Bi-Te}. They are closely related to the structure of the nonmagnetic TI Bi$_2$Te$_3$ (Fig.~\ref{fig:structure_Bi2Te3}). MnBi$_2$Te$_4$ is built up of septuple (ST) Te--Bi--Te-Mn--Te--Bi--Te layers stacked along the $c$-axis. The ST layer differs from the quintuple (QT) Te--Bi--Te--Bi--Te layer of Bi$_2$Te$_3$ by the presence of an additional Mn--Te atomic sublayer. In MnBi$_4$Te$_7$ magnetic ST layers are separated by nonmagnetic QT layers. MnBi$_2$Te$_4$ and MnBi$_4$Te$_7$ order antiferromagnetically at $T_{\rm N} = 24$\,K and 13\,K, respectively \cite{Otrokov2019,Vidal2019}. The magnetic moments of the Mn$^{2+}$ ($S = 5/2$) ions in the ST layers order out-of-plane ferromagnetically while the intelayer coupling between the QT layers is antiferromagnetic. Both compounds are self-doped magnetic semiconductors and feature topological surface electronic states \cite{Otrokov2019,Vidal2019}.

The spin dynamics in the AFM-ordered and paramagnetic states of MnBi$_2$Te$_4$ and MnBi$_4$Te$_7$ and its possible impact on the surface electronic states was addressed by the AFMR/ESR spectroscopy in Refs.~\cite{Otrokov2019,Vidal2019,Alfonsov2021,Alfonsov2021a}. Multifrequency ESR measurements in the range 75--330\,GHz made it possible to map the AFMR excitation branches in the ordered state   and study their evolution by raising the temperature above $T_{\rm N}$ \cite{Vidal2019,Alfonsov2021a}.   
\begin{figure}
	\centering
	\includegraphics[width=0.65\linewidth]{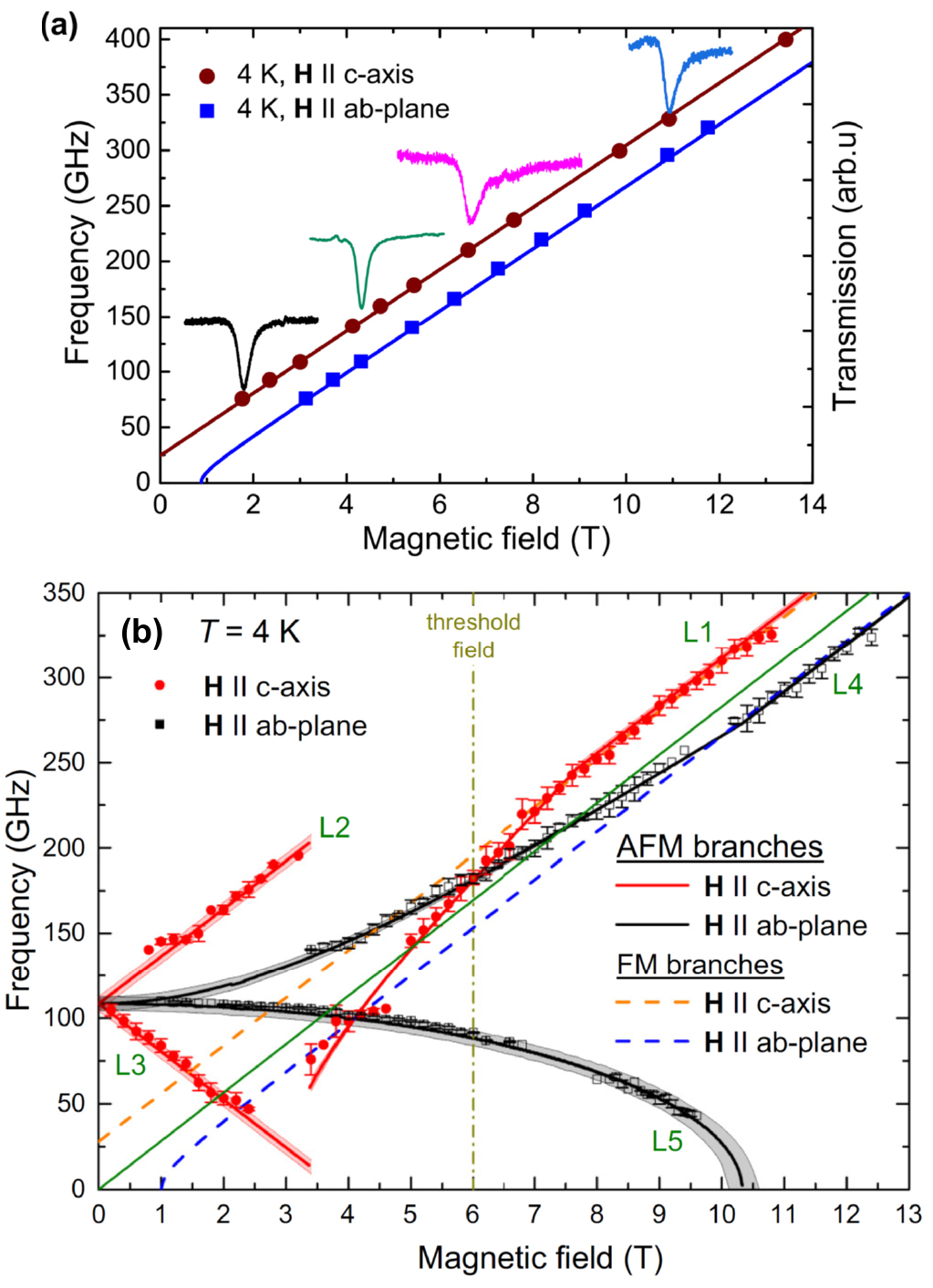}
	\caption{(a) Resonance branches $\nu(H)$ at $T = 4$\,K\,$\ll T_{\rm N}$ for MnBi$_4$Te$_7$ (a) and for MnBi$_2$Te$_4$ (b). Typical absorption signals of the  MnBi$_4$Te$_7$ compound at selected frequencies are shown in panel (a). In both panels symbols are experimental data and solid and dashed lines are results of theoretical modeling. The green solid line in panel (b) denotes the paramagnetic resonance branch $\nu = (g\mu_{\rm B}/h)H$. 
		(Reprinted with permission from A.~Alfonsov {\it et al.}, Phys. Rev. B. {\bf 104} (19), 195139 (2021) \cite{Alfonsov2021}. Copyright (2021) by the American Physical Society.)} 
	\label{fig:AFMR_Mn-Bi-Te}
\end{figure}

Although MnBi$_4$Te$_7$ orders AFM at $T_{\rm N} =13$\,K, the interlayer coupling between magnetic ST layers separated by nonmagnetic QT layers is such weak that an application of a small magnetic field of the order of some tens of mT polarizes the Mn spins and turns the system into the forced FM state. In this regime MnBi$_4$Te$_7$ demonstrates the resonance response typical for an easy-axis ferromagnet [Fig.~\ref{fig:AFMR_Mn-Bi-Te}(a)] for which analytical solutions do exist \cite{Turov}:
\begin{eqnarray}
		& {\rm  H\parallel c-axis:}   & h \nu = \text{g} \mu_B \mu_{0} (H+|H_a|)  \label{eq:EAHIIc}\\
		&{\rm  H\parallel ab-plane:}   & h \nu = \text{g} \mu_B \mu_{0} \sqrt{H(H-|H_a|)}\ . \label{eq:EAHIIab}
\end{eqnarray}
Fitting the data to Eqs.~(\ref{eq:EAHIIc}) and (\ref{eq:EAHIIab}) yields the value of the total anisotropy field $H_a$. After a proper subtraction from $H_a$ of the demagnetizing field due to the platelet-like shape of the sample the intrinsic magnetocrystaline energy of MnBi$_4$Te$_7$ was evaluated to be $E_{\rm a} = -0.18 \pm 0.01$\,meV per Mn in one magnetic ST layer.

The AFM excitation spectrum of MnBi$_2$Te$_4$ is more complex [Fig.~\ref{fig:AFMR_Mn-Bi-Te}(b)]. At fields smaller than $\sim 5$\,T the resonance branches are typical for a uniaxial easy-axis antiferromagnet. For $\mathbf{H}$ along the easy axis ($c$-axis) there are two branches, one ascending (L2) and one descending (L3),  that collapse at the spin-flop field $\mu_{0}H_{\rm sf} = 3.5$\,T, above which a new ascending resonance branch L1 appears. For $\mathbf{H}$ in the hard plane ($ab$-plane) there are two non-linear dispersing modes L4 and L5. Surprisingly, the L1 and L3 branches cross each other at a threshold field of $\sim 6$\,T above which they follow a linear field dependence. This observation is at odds with the canonical spin-wave theory according to which the branches L1 and L4 should asymptotically approach the paramagnetic line $\nu = (g\mu_{\rm B}/h)H$ but never cross it under the condition that the AFM exchange energy $E_{\rm exch}$ is much larger than the magnetic anisotropy energy $E_{\rm a}$ \cite{Turov}. This puzzle was solved by numerical modeling of the branches on the basis of a linear spin-wave theory with the second quantization formalism \cite{Turov,Holstein1940}. An excellent agreement with experiment was achieved [solid red and black lines in Fig.~\ref{fig:AFMR_Mn-Bi-Te}(b)]  yielding the AFM interplane exchange energy $E_{\rm exch} = 0.65 \pm 0.01$\,meV per Mn in one ST layer and $E_{\rm a} = -0.21 \pm 0.01$\,meV per Mn in one ST layer. The unusual crossing of the L1 and L4 branches appeared to be due to the similar strength of $E_{\rm exch}$ and $E_{\rm a}$ in MnBi$_2$Te$_4$ and the closeness of these energy scales to that of the applied magnetic field. Thus, such a strong applied magnetic field can completely decouple the magnetic ST layers and turn the formerly AFMR modes L1 and L4 into the linear dispersing FM excitations according to Eqs.~(\ref{eq:EAHIIc}) and (\ref{eq:EAHIIab}) [dashed red and black straight lines in Fig.~\ref{fig:AFMR_Mn-Bi-Te}(b)]. The similarity of the anisotropy energies of MnBi$_2$Te$_4$ and MnBi$_4$Te$_7$  suggests that it is an intrinsic property of the magnetic ST layers little affected by the insertion of nonmagnetic QT layers.

Temperature dependent measurements  of the resonance signals evidence the 2D character of the FM in-plane correlations in the weakly coupled magnetic ST layers in MnBi$_4$Te$_7$ persisting far above the 3D ordering temperature $T_{\rm N} = 13$\,K. Specifically, the zero field excitation gap of the FM modes does not close at $T_{\rm N}$ but vanishes only at $\sim 30$\,K. The resonance field deviates from its isotropic paramagnetic value up to the same temperature. In contrast, the zero field excitation gap for MnBi$_2$Te$_4$ closes at $T_{N} = 24$\,K and the resonance field of the signals measured at small applied fields rapidly approaches the isotropic paramagnetic value above $T_{\rm N}$ as expected for a 3D magnet with significant inter-layer coupling. However, measurements at fields stronger than the threshold field of $\sim 6$\,T, i.e., in the regime where the magnetic ST layers become decoupled, the 2D FM correlations remain prominent up to temperatures of $\sim 2T_{\rm N}$ manifesting in the anisotropic shifts of the resonance signals. 
\begin{figure}
	\centering
	\includegraphics[width=\linewidth]{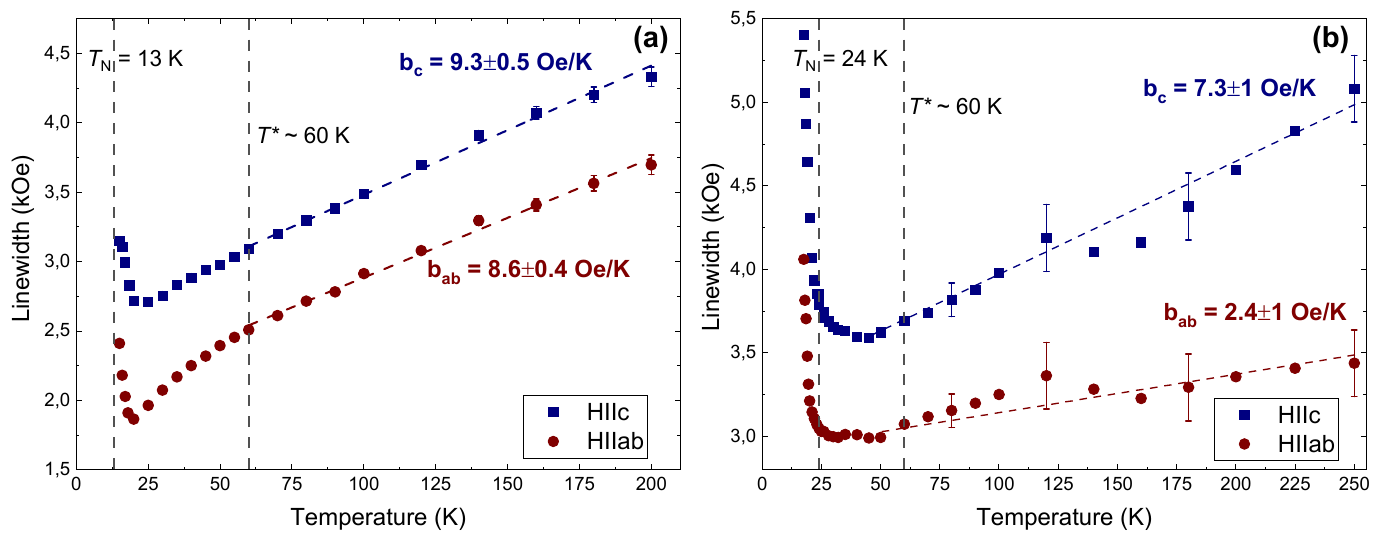}
	\caption{Temperature dependence of the ESR linewidth at $\nu = 9.56$\,GHz and at two orientations of the applied magnetic field for MnBi$_4$Te$_7$ (a) and for MnBi$_2$Te$_4$ (b).
	Dashed lines are the fits to the Korringa dependence $\Delta(H) = \Delta H_0 + bT$. Vertical dashed lines indicate the magnetic ordering temperatures $T_{\rm N}$ and the onset of the Korringa relaxation regime $T^\ast$. 
	(Reprinted with permission from A.~Alfonsov {\it et al.}, Phys. Rev. B. {\bf 103} (18), 180403 (2021) \cite{Alfonsov2021a}. Copyright (2021) by the American Physical Society.)} 
	\label{fig:Korringa_Mn-Bi-Te}
\end{figure}

The dynamics of the Mn spins in the paramagnetic state of MnBi$_2$Te$_4$ and MnBi$_4$Te$_7$ and its possible  relation to the surface electronic structure was studied with the X-band ESR spectroscopy in Refs.~\cite{Otrokov2019,Alfonsov2021a}. Since both compounds are doped semiconductors demonstrating bulk metallic conductivity one expected the Korringa mechanism to be the main channel of relaxation of Mn spins similar to, e.g., the Mn-doped  Bi$_2$Te$_3$ (see, Sect.~\ref{subsec:Mag_doped_vdW_TI}). Indeed, upon increasing the temperature above $T_{\rm N}$ the critical broadening of the ESR line due to the short-range FM correlations ceases and the linear in temperature Korringa relaxation regime characteristic of an uncorrelated metallic paramagnet sets in at $T^\ast \sim 60$\,K (Fig.~\ref{fig:Korringa_Mn-Bi-Te}). The data above $T^\ast$ was fitted to the Korringa dependence $\Delta(H) = \Delta H_0 + bT$, where $\Delta H_0$ is the residual linewidth and $b$ is the Korringa slope defined in Eq.~(\ref{eq:KorringaSlope}). In most of the metallic compounds the parameter $b$ was found to be independent of the direction of the applied field. Evidently, that was also the case for MnBi$_4$Te$_7$ where the Korringa slope was almost the same for both field orientations with $b_{\rm c}\sim b_{\rm ab} \sim 9$\,Oe/K  [Fig.~\ref{fig:Korringa_Mn-Bi-Te}(a)]. However, the Korringa slope for MnBi$_2$Te$_4$ was found to be surprisingly anisotropic with $b_{\rm c} = 7$\,Oe/K$\, \gg b_{\rm ab} = 2.4$\,Oe/K [Fig.~\ref{fig:Korringa_Mn-Bi-Te}(b)]. 

To understand this striking observation the authors have carried out density-functional-theory (DFT) calculations of the bulk electronic structure. The results show that the density of states of the bulk charge carriers in a certain range of the carrier concentrations depends on the direction of the Mn magnetic moments. This non-trivial property causes the anisotropy of the dynamics of the localized Mn moments reflected in the anisotropy of the Korringa slope $b$ which, due to the specific details of the band structure, is very strong in MnBi$_2$Te$_4$ and is much less pronounced in MnBi$_4$Te$_7$. This finding helped to find an explanation of the unexpected temperature evolution of the band structure of topological surface states observed in MnBi$_2$Te$_4$ by angle-resolved photoemission spectroscopy (ARPES) \cite{Otrokov2019}. The energy gap opened at the surface Dirac point at $T < T_{\rm N}$ due to the static AFM order of the Mn spins did not close completely at $T_{\rm N}$ and persisted up to much higher temperatures. In view of the ESR results this was ascribed to the strong anisotropy of the Mn spin fluctuations caused by their anisotropic relaxation on bulk charge carriers that produces a nonzero instantaneous out-of-plane magnetic field acting on the surface electronic states on a timescale much longer than that in the ARPES experiment which then detects a kind of a dynamically opened gap at the Dirac point. 

The sensitivity of the density of states to the particular orientation of the Mn moments is obviously related, among other factors,  to the strong spin-orbit coupling in particular due to the presence of the heavy element Bi in MnBi$_2$Te$_4$. Indeed, recent ESR experiments in Ref.~\cite{Sahoo2023} on the new candidate material for ferromagnetic topological insulator MnSb$_2$Te$_4$ with mich lighter Sb instead of Bi  documented a completely isotropic Korringa behavior of the Mn ESR linewidth and the DFT calculations showed that the band structure is insensitive to the orientation of the Mn moments.

\section{Conclusion}\label{sec13}
In conclusion, this article summarized the achievements  of the ESR, AFMR and FMR studies on quasi-2D van der Waals magnets in the elucidation of important fine details of the magnetically ordered ground state of these compounds, in particular a precise determination of the type, the symmetry  and the magnitude of the intrinsic magnetocrystalline anisotropy. Also, these investigations provided comprehensive insights into the low-energy magnon excitations in the ordered phase and the details of the correlated spin dynamics in the paramagnetic state. Owing to the very weak interlayer magnetic coupling 
%
%
many of these materials demonstrate in the bulk form signatures of intrinsic 2D behavior both in the static and dynamic sectors. Therefore, the obtained results on the studied bulk single crystals enable  reliable predictions for the monolayers of these materials in particular in terms of the stability of the long-range magnetic order and the damping of the spin wave excitations, which are the decisive criteria for the use of monolayers in the spin-electronic devices. Furthermore, the ESR spectroscopy contributed to the significant progress in the understanding of the interplay between magnetism and electronic structure in semiconducting and metallic van der Waals compounds which affects the magnetic anisotropy and thus the type of magnetic order and also gives rise to novel magneto-electronic effects which are realized, e.g., in  magnetic topological insulators.

\backmatter

\bmhead{Acknowledgments}

The authors gratefully acknowledge financial support by the Deutsche Forschungsgemeinschaft (DFG) through Projects No. 447482487, 499461434, 455319354, and within the Collaborative Research Center SFB 1143 ``Correlated Magnetism - From Frustration to Topology'' (project-id 247310070), and the Dresden-W\"urzburg Cluster of Excellence (EXC 2147) ``ct.qmat - Complexity and Topology in Quantum Matter'' (project-id 390858490).

\section*{Declarations}

\bmhead{Funding}
\

\noindent
Deutsche Forschungsgemeinschaft, projects 447482487, 499461434, 455319354, 247310070, and 390858490.

\bmhead{Conflict of interest}
\

\noindent
\

\noindent
The authors declare no conflict of interest.

\bmhead{Ethics approval}
\

\noindent
not applicable

\bmhead{Availability of data and materials}
\

\noindent
not applicable

\bmhead{Authors' contributions}
\

\noindent
All authors jointly developed the conception of the review. V.K. wrote the first draft of the manuscript with active participation of A.A. and B.B. All authors edited its final version.

%
%
%
%


\bibliography{search_ESR_vanderWaals,search_ESR_topins,general}

\end{document}